\documentclass{article}

 \usepackage[preprint]{neurips_2026}

\usepackage[utf8]{inputenc} % allow utf-8 input
\usepackage[T1]{fontenc}    % use 8-bit T1 fonts
\usepackage{hyperref}       % hyperlinks
\usepackage{url}            % simple URL typesetting
\usepackage{booktabs}       % professional-quality tables
\usepackage{amsfonts}       % blackboard math symbols
\usepackage{nicefrac}       % compact symbols for 1/2, etc.
\usepackage{microtype}      % microtypography
\usepackage{xcolor}         % colors

\usepackage{amsmath,amsfonts,bm}

\def\eqref#1{equation~\ref{#1}}
\def\1{\bm{1}}

\DeclareMathAlphabet{\mathsfit}{\encodingdefault}{\sfdefault}{m}{sl}
\SetMathAlphabet{\mathsfit}{bold}{\encodingdefault}{\sfdefault}{bx}{n}

\usepackage{colortbl}
\usepackage{arydshln}
\usepackage{enumitem}
\usepackage{multirow}
\usepackage{tcolorbox}
\usepackage{svg}
\usepackage{listings}
\usepackage{inconsolata}
\usepackage{xspace}
\usepackage{graphicx}
\usepackage{caption}
\usepackage{float}
\usepackage{adjustbox}
\usepackage{alltt}
\usepackage{lineno}
\usepackage{algorithm}
\usepackage{algpseudocode}
\usepackage[normalem]{ulem}
\usepackage{tcolorbox}
\tcbuselibrary{skins,breakable}

\tcbset{
  aibox/.style={
    width=400.18663pt,
    top=10pt,
    colback=black!05,
    colframe=black!20,
    colbacktitle=black!50,
    center,
  }
}

\tcbset{
  aiboxbreakable/.style={
    width=400.18663pt,
    top=10pt,
    colback=black!05,
    colframe=black!20,
    colbacktitle=black!50,
    center,
    breakable,
  }
}

\newtcolorbox{AIBox}[2][]{aibox,title=#2,#1}
\newtcolorbox{AIBoxBreak}[2][]{aiboxbreakable,title=#2,#1}

\tcbset{
  aibox/.style={
    top=8pt,
    bottom=3pt,
    colback=white,
    colframe=black,
    colbacktitle=black,
    enhanced,
    center,
    attach boxed title to top left={yshift=-0.1in,xshift=0.15in},
    boxed title style={boxrule=0pt,colframe=white,},
  }
}
\newtcolorbox{AIbox}[3][]{aibox, width=#2, title=#3,#1}

\DeclareCaptionType{prompt}[Prompt][List of Prompts]
\DeclareCaptionType{artifact}[Artifact][List of Artifacts]
\DeclareCaptionType{example}[Example][List of Example]

\newcommand{\eg}{\textit{e.g.}\xspace}

\renewcommand{\cite}{\citep}

\newcommand\openai{\raisebox{-3pt}{\includegraphics[height=1.6em]{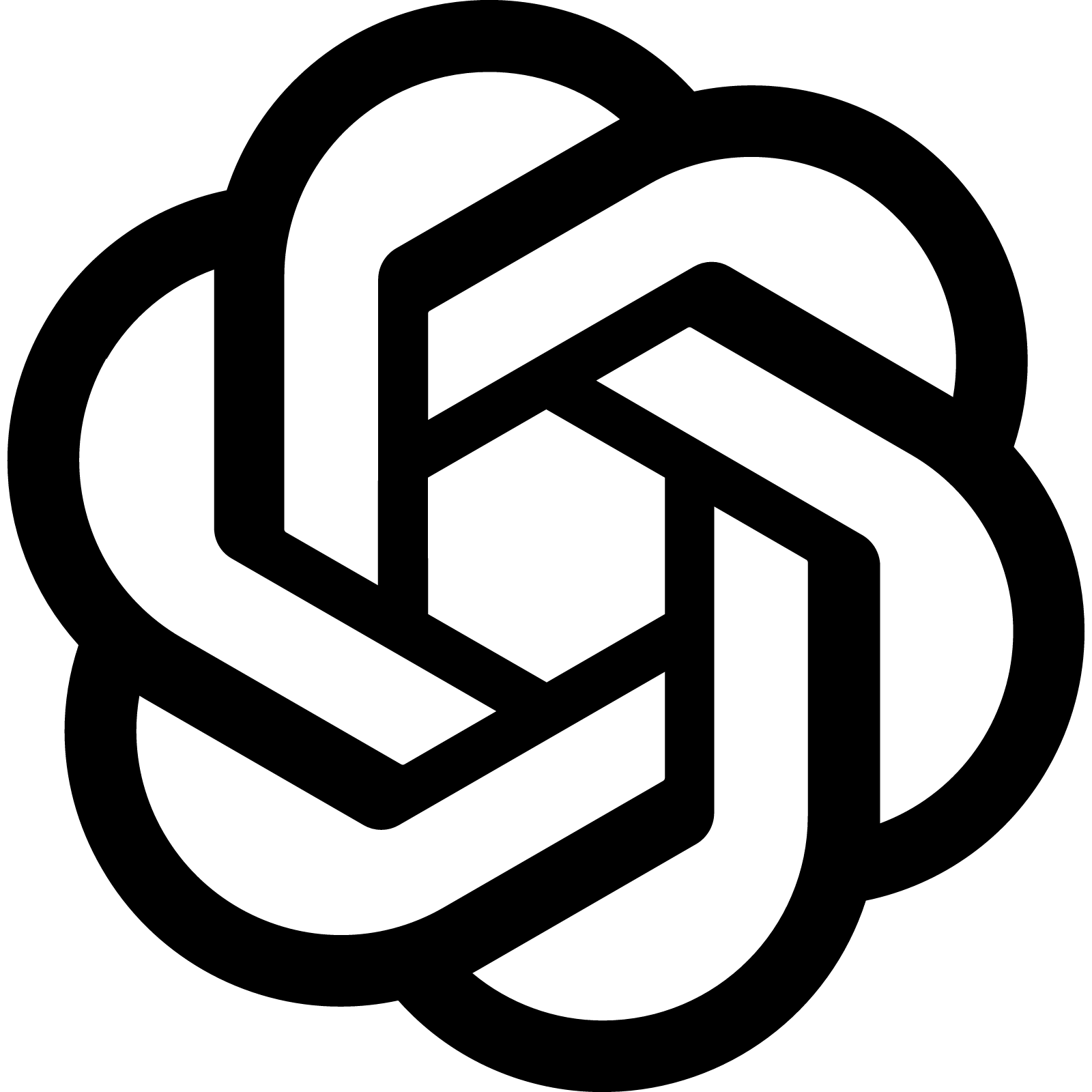}}}
\newcommand\anthropic{\raisebox{-3pt}{\includegraphics[height=1.6em]{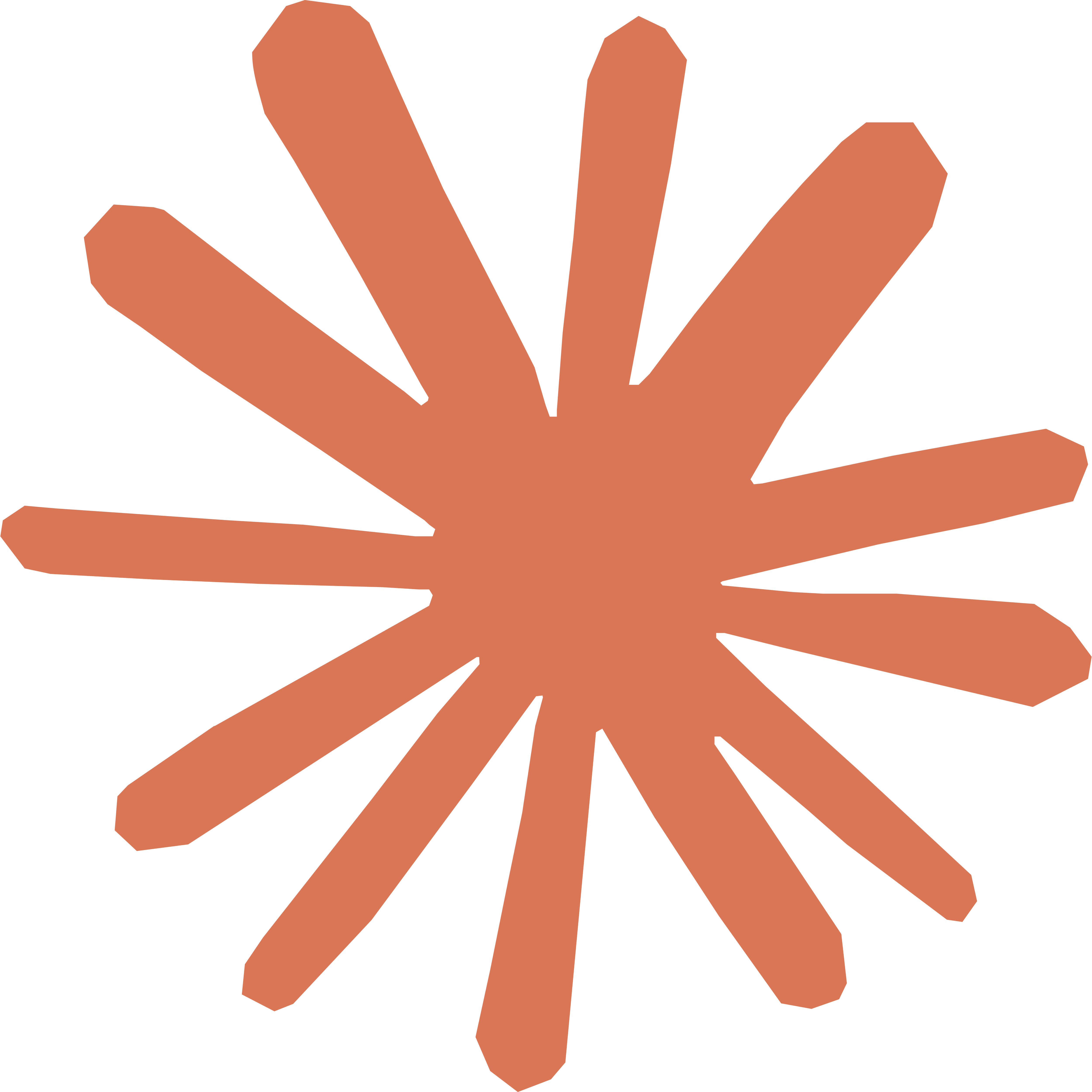}}}
\newcommand\minimax{\raisebox{-3pt}{\includegraphics[height=1.6em]{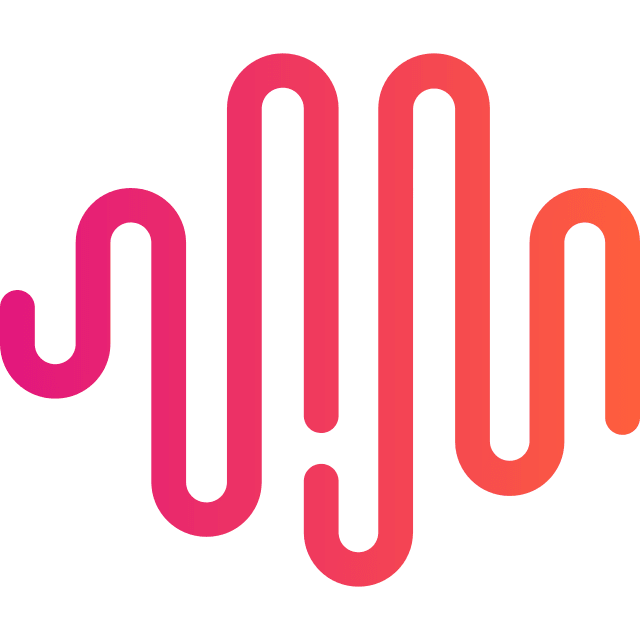}}}
\newcommand\zai{\raisebox{-3pt}{\includegraphics[height=1.6em]{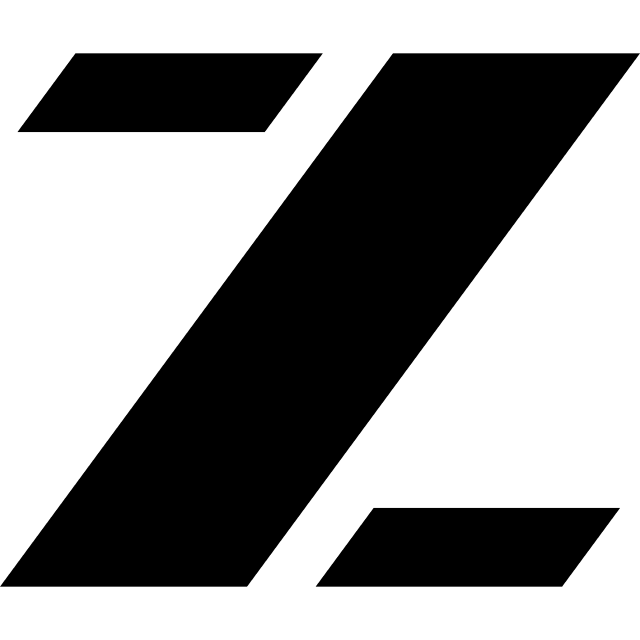}}}
\newcommand\claudecode{\raisebox{-3pt}{\includegraphics[height=1.6em]{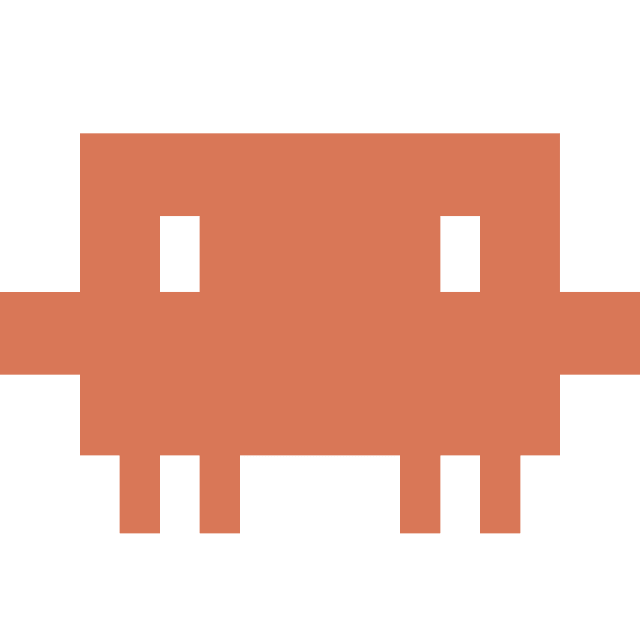}}}
\newcommand\codex{\raisebox{-3pt}{\includegraphics[height=1.6em]{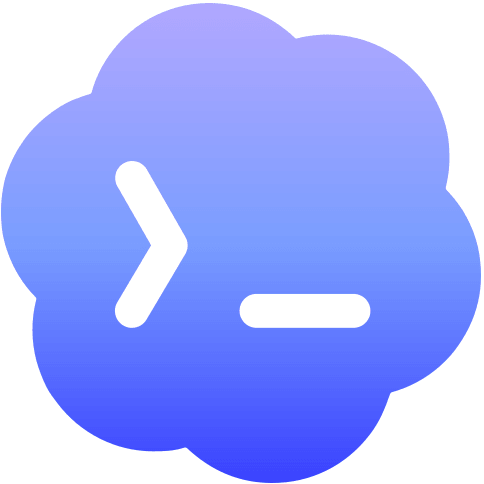}}}
\newcommand\opencode{\raisebox{-3pt}{\includegraphics[height=1.6em]{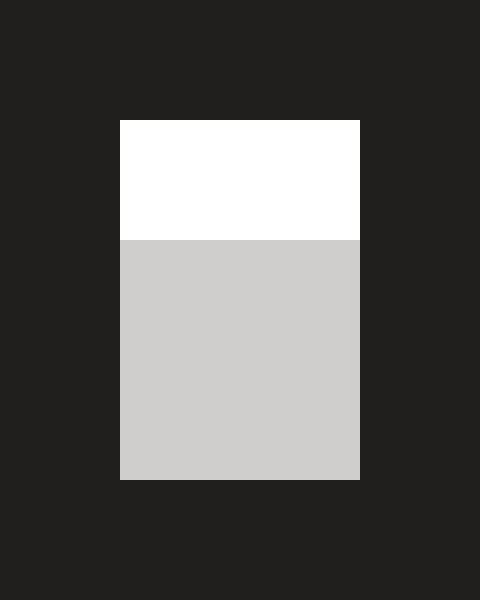}}}

\newcommand{\methodname}{\textsc{\textbf{SWE-Chain}}\xspace}
\newcommand{\synthesis}{{\textbf{DecompSynth}}\xspace}
\title{\methodname: Benchmarking Coding Agents on Chained Release-Level Package Upgrades}

\author{
Man Ho Lam$^{1}$ \quad
Chaozheng Wang$^{1}$\thanks{Corresponding author.} \quad
Hange Liu$^{2}$ \quad
Jingyu Xiao$^{1}$ \quad
Haau{-}sing Li$^{3,4}$ \\ \bf 
Jen{-}tse Huang$^{5}$ \quad
Terry Yue Zhuo$^{6}$ \quad
Michael R. Lyu$^{1}$ \\
$^1$The Chinese University of Hong Kong \quad\quad $^2$Independent \quad\quad $^3$ELLIS \\
$^4$Technical University of Darmstadt \quad\quad $^5$Johns Hopkins University \quad\quad $^6$Monash University \\
\texttt{mhlam@link.cuhk.edu.hk} \quad\quad \texttt{czwang23@cse.cuhk.edu.hk}
}

\begin{document}

\maketitle

\begin{center}
\vspace{-20pt}
\href{https://github.com/CUHK-ARISE/SWE-Chain}{\includegraphics[width=10pt]{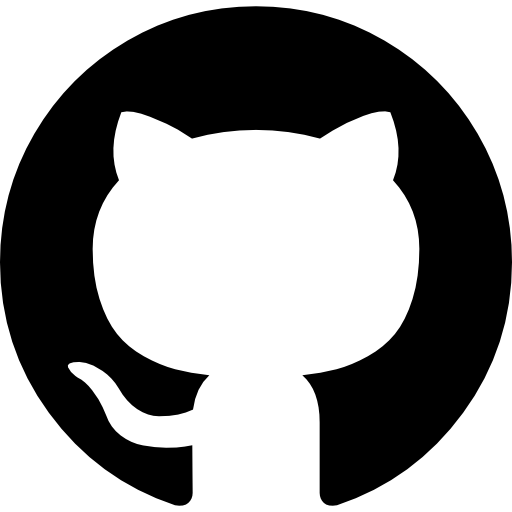}\ \textbf{GitHub Code}} \quad \quad
\href{https://huggingface.co/datasets/mhlam/SWE-Chain}{\includegraphics[width=10pt]{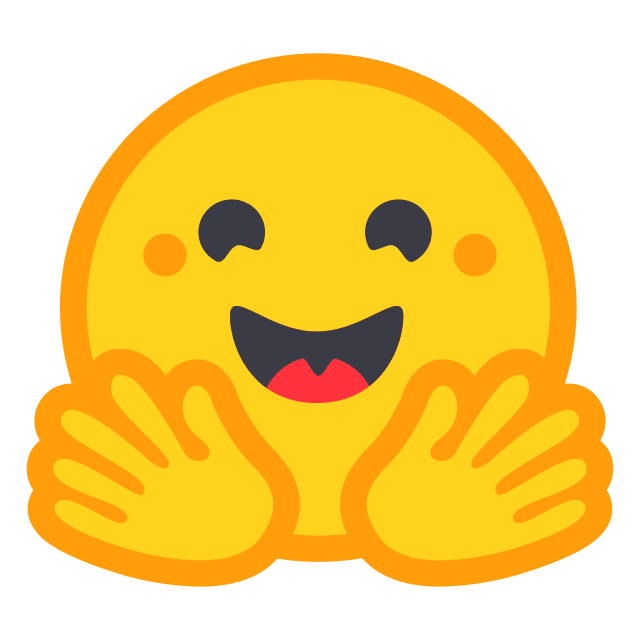}\ \textbf{HuggingFace Dataset}}
\vspace{10pt}
\end{center}

\begin{abstract}
Coding agents powered by large language models are increasingly expected to perform realistic software maintenance tasks beyond isolated issue resolution.
Existing benchmarks have shifted toward realistic software evolution, but they rarely capture continuous maintenance at the granularity of package releases, where changes are bundled, shipped, and inherited by subsequent versions.
We present \methodname, a benchmark for evaluating agents on \emph{chained release-level package upgrades}, where each transition builds on the agent's prior codebase.
To produce upgrade specifications, we design a \emph{divide-and-conquer synthesis pipeline} that aligns release notes with code diffs for each version transition, ensuring the requirements are grounded in actual code changes, informative to agents, and feasible to implement.
\methodname contains \textbf{12} upgrade chains across 9 real Python packages, with \textbf{155} version transitions and 1,660 grounded upgrade requirements.
Across nine frontier agent-model configurations, agents achieve an average of \textbf{44.8\%} resolving, 65.4\% precision, and 50.2\% F1 under the Build+Fix regime, with Claude-Opus-4.7 (Claude Code) leading at \textbf{60.8\%} resolving, 80.6\% precision, and 68.5\% F1.
These results show that \methodname is both feasible and discriminative, and reveal that current agents still struggle to make correct upgrades across chained package releases without breaking existing functionality.

\end{abstract}

\addtocontents{toc}{\protect\setcounter{tocdepth}{-1}}
\section{Introduction}
\label{sec:introduction}

Powered by advanced Large Language Models (LLMs), coding agents such as Codex~\cite{openai2025codex} and Claude Code~\cite{anthropic2025claudecode} have evolved into software engineering assistants that can navigate repositories, execute commands, inspect errors, and edit code through tool-use interfaces~\cite{yang2024sweagent, wang2025openhands, xie2025swefixer, xia2025demystifying}.
Accordingly, benchmarks have evolved from task-level problems~\cite{liu2023evalplus, gu2024cruxeval, jain2025livecodebench, zhuo2025bigcodebench, lam2025codecrash} to repository-level issue resolution in real-world projects~\cite{jimenez2024swebench, liu2024repobench, li2024evocodebench, deng2025swebenchpro}.
Complementary efforts further broaden software-engineering evaluation to multimodal settings involving visual artifacts~\cite{yang2024swebenchmultimodal, wan2024mrweb, xiao2024interaction2code, xiao2025designbench}.
More recently, benchmarks like $\tau$-Bench~\cite{yao2025taubench}, Terminal-Bench~\cite{merrill2026terminalbench}, and AgencyBench~\cite{li2026agencybench} have shifted towards realistic software development scenarios with tool use, terminal interaction, and multi-turn decision-making.

Consequently, growing attention has shifted towards long-horizon software engineering that goes beyond isolated issue resolution, raising a fundamental design question: how should long software trajectories be segmented and evaluated?
SWE-CI~\cite{chen2026sweci} defines tasks by benchmark-selected base-to-oracle commit pairs, SlopCodeBench~\cite{orlanski2026slopcodebench} uses synthetic checkpoints, and EvoClaw~\cite{deng2026evoclaw} reconstructs mid-release milestones.
These scenarios provide important insights into modeling software engineering lifecycles, exposing failures in iterative code degradation and in continuous integration (CI)-style maintainability. However, none of these segmentation boundaries corresponds to a moment when an upstream maintainer actually ships a software version.
Specifically, they capture merely part of the real-world package maintenance dynamics, where changes are bundled at release boundaries, and earlier mistakes can carry over to later releases.

SWE-EVO~\cite{thai2026sweevo} takes the first step in this direction by treating release versions as evaluation units.
However, its specifications are constructed by concatenating release notes with raw content from GitHub issues and pull requests, leaving agents to process noisy artifacts such as terminal dumps, reproduction snippets, and image references rather than unified and coherent upgrade specifications.
Furthermore, each task is an isolated end-to-end release transition, which does not evaluate whether agents can continuously maintain a codebase built on their previous modifications.
This motivates our central research question:
\begin{center}
\emph{Can agents maintain a real package through a chain of releases while carrying their own changes forward without breaking existing functionality?}
\end{center}

We therefore introduce \textbf{\methodname}, a benchmark for evaluating coding agents on \textit{chains of consecutive release upgrades}.
To address the noisy-specification issue of prior release-level evaluation, we introduce \synthesis, a \textit{divide-and-conquer synthesis pipeline} that synthesizes specifications from maintainer-authored release notes and actual behavioral changes in the gold code diff, ensuring that the requirements in the specification are \textbf{grounded} in actual code changes, \textbf{informative} to the agent, and \textbf{feasible} to implement.
The version-aligned specifications allow us to compose individual version transitions into upgrade chains, in which the agent-produced codebase at one release becomes the starting point for the next, enabling us to model the release-level maintenance cycle.
Our contributions include:
\begin{enumerate}[leftmargin=*]
\item \textbf{\methodname}: We introduce the first benchmark for chained release-level maintenance, comprising 12 upgrade chains across 9 packages, 155 version transitions, and 1,660 grounded tasks.
\item \textbf{\synthesis}: We implement and open-source a lightweight yet scalable pipeline that leverages release notes and code diffs to construct high-quality upgrade specifications, replacing manual authoring and noisy one-shot synthesis.
\item \textbf{Comprehensive evaluation}: 
We evaluate 9 frontier model-agent configurations, achieving an average resolving rate of $\mathbf{44.8\%}$, precision of $\mathbf{65.4\%}$, and F1 of $\mathbf{50.2\%}$ under the Build+Fix regime, which allows one fix attempt for execution-level errors.
These results show current agents can partially solve chained upgrades but still struggle to preserve functionality across releases.
\end{enumerate}
\section{\methodname}
\label{sec:methodology}
\begin{figure}[t]
    \centering
    \includegraphics[width=1.0\linewidth]{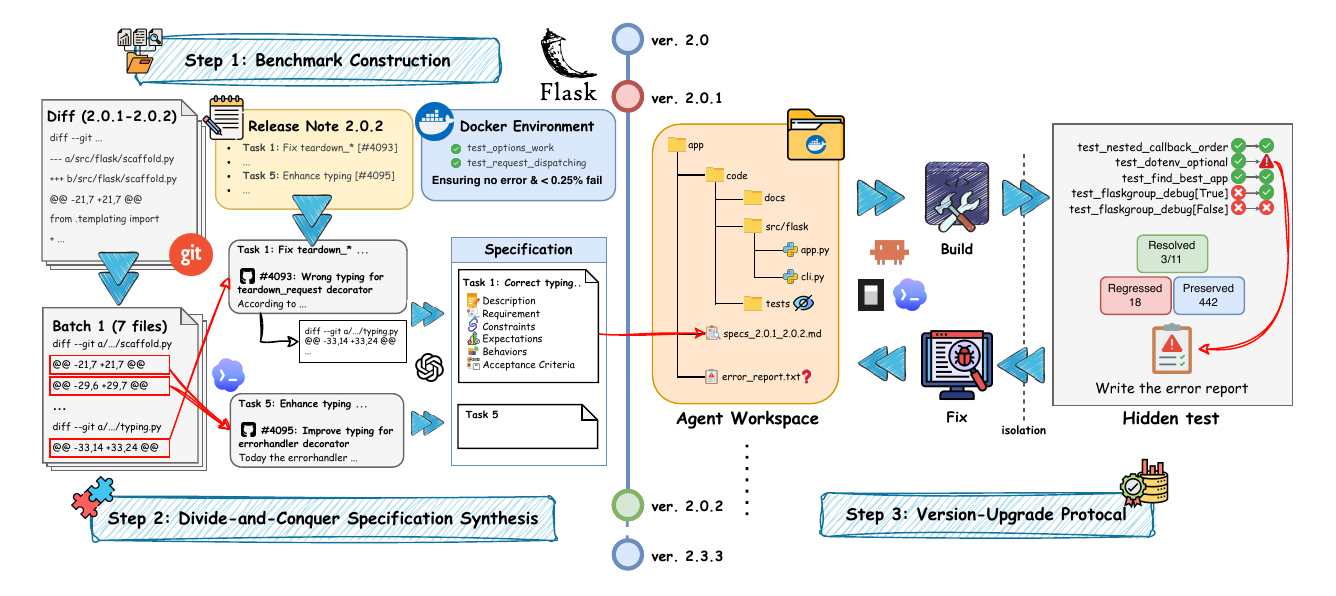}
    \caption{Overview of the \methodname pipeline. For each version upgrade, we collect release notes, extract diffs, and prepare environments. An agent matches hunks to tasks and synthesizes specifications, which are used to evaluate agents sequentially, with an extra fixing step for errors.}
    \label{fig:overview}
\end{figure}

This section outlines the construction of \methodname, including data collection, environment setup, cross-version validation, specification synthesis, and evaluation protocol, as illustrated in Figure~\ref{fig:overview}.

\begin{table*}[t]
\centering
\small
\setlength{\tabcolsep}{4pt}
\renewcommand{\arraystretch}{1.15}
\caption{Scale of the chains in \methodname. LOCD is the total Python code diff lines across all upgrade steps, and LOC$_{\text{start}}^{\text{no tests}}$ counts lines in the starting codebase excluding the test suite. Struck-through versions denote skipped releases (\eg, yanked versions or negligible upgrades).}
\label{tab:chain_overview}
\begin{tabular}{llcccc}
\toprule
\textbf{Package} & \textbf{Chain} & \textbf{\#Upgrades} & \textbf{\#Tasks (+Extra)} & \textbf{LOCD} & \textbf{LOC$_{\text{start}}^{\text{no tests}}$} \\
\midrule
\href{https://pypi.org/project/attrs/}{attrs} & 21.3.0 $\rightarrow$ 26.1.0 & $13$ & $92(+6)$ & $7,834$ & $13\text{k}$ \\
\multirow[c]{2}{*}{\href{https://pypi.org/project/conan/}{conan}} & 2.12.0 $\rightarrow$ 2.20.1 & $16$ & $243(+18)$ & $13,901$ & $67\text{k}$ \\
 & 2.23.0 $\rightarrow$ 2.28.1 & $11$ & $167(+6)$ & $7,284$ & $76\text{k}$ \\
\href{https://pypi.org/project/Flask/}{Flask} & 2.0.0 $\rightarrow$ 2.3.3 & $17$ & $107(+4)$ & $8,013$ & $29\text{k}$ \\
\href{https://pypi.org/project/Jinja2/}{Jinja2} & 2.8 $\rightarrow$ 2.10.3 & $12$ & $67(+7)$ & $4,730$ & $23\text{k}$ \\
\href{https://pypi.org/project/poetry/}{poetry} & 1.5.0 $\rightarrow$ 1.8.5 (\sout{1.6.1}) & $10$ & $196(+3)$ & $8,977$ & $21\text{k}$ \\
\href{https://pypi.org/project/PyJWT/}{PyJWT} & 2.0.0 $\rightarrow$ 2.12.1 & $15$ & $145(+5)$ & $3,369$ & $3.4\text{k}$ \\
\multirow[c]{2}{*}{\href{https://pypi.org/project/pytest/}{pytest}} & 7.0.0 $\rightarrow$ 7.4.4 & $16$ & $134(+6)$ & $7,225$ & $79\text{k}$ \\
 & 8.0.0 $\rightarrow$ 8.3.5 (\sout{8.1.0, 8.3.1}) & $12$ & $114(+7)$ & $10,328$ & $89\text{k}$ \\
\href{https://pypi.org/project/urllib3/}{urllib3} & 2.0.7 $\rightarrow$ 2.6.3 & $12$ & $72(+7)$ & $8,141$ & $17\text{k}$ \\
\multirow[c]{2}{*}{\href{https://pypi.org/project/xarray/}{xarray}} & 2022.11.0 $\rightarrow$ 2023.7.0 & $10$ & $87(+7)$ & $13,082$ & $119\text{k}$ \\
 & 2025.6.0 $\rightarrow$ 2026.2.0 & $11$ & $141(+19)$ & $14,120$ & $152\text{k}$ \\
\bottomrule
\end{tabular}
\vspace{-0.5em}
\end{table*}

\subsection{Benchmark Construction}
\label{sec:met:benchmark}

We collect released package versions from PyPI or GitHub and retrieve the corresponding release notes, as well as relevant GitHub issues and pull requests (PRs).
Table~\ref{tab:chain_overview} summarizes the selected upgrade chains and their scale, with more details in Appendix~\ref{app:scale}.
We then implement lightweight, package-specific fetchers and decomposition functions to automatically parse release notes into raw tasks, followed by a manual quality review to normalize the formatting and exclude unsuitable releases or chains.
To ground each version upgrade in implementation evidence, we extract the Python code and test diffs between consecutive versions using \texttt{git diff}, which we then use to synthesize the specification in \S\ref{sec:met:synthesis}.

\paragraph{Docker environment setup.}
To provide a consistent chain-level Docker environment, we manually design a Dockerfile for each chain.
Each Dockerfile pins the Python interpreter and cross-version dependencies, while allowing other packages to resolve to their release-specific versions.
This enables a single image to run tests for all versions while preserving each release's dependency landscape.
The sanity check before cross-version validation ensures that every version executes \textbf{without setup errors} and yields \textbf{fewer than 0.25\% failed test cases} on its corresponding test suite.

\paragraph{Cross-version validation.}
For a version upgrade $\mathcal{V}_i$ from $v_{i-1}$ to $v_i$, we identify upgrade-related test cases by running the original test suite $\mathcal{Q}_{v_i}$ on both the old and new gold codebases, $c_{v_{i-1}}$ and $c_{v_i}$, respectively.
Following SWE-Bench~\cite{jimenez2024swebench}, we include both the \texttt{FAIL\_TO\_PASS} and \texttt{ERROR\_TO\_PASS} cases as upgrade-related tests.
Formally, let $r(q,c) \in \{\mathrm{pass}, \mathrm{fail}, \mathrm{error}\}$ denote the execution result of test case $q$ on codebase $c$. We define the upgrade-related test set as
\begin{equation}
\mathcal{U}_i =
\{q \in \mathcal{Q}_{v_i}
\mid r(q,c_{v_{i-1}}) \in \{\mathrm{fail}, \mathrm{error}\}
\land r(q,c_{v_i}) = \mathrm{pass}\}.
\label{eq:upgrade-related}
\end{equation}

\subsection{\synthesis}
\label{sec:met:synthesis}
We synthesize the specification for each consecutive version pair primarily from two sources:
(i) \textbf{Release note} $\mathcal{R}$ with corresponding fetched GitHub issues or pull requests, which is represented as a list of tasks $[\mathcal{T}^{(1)}, \ldots, \mathcal{T}^{(n)}]$; and
(ii) \textbf{Diff context} $\mathcal{D}$, consisting of source-code diffs as the primary grounding evidence and test-suite diffs for calibrating behavioral intent.
It is organized as a list of file-level diffs $[\mathcal{D}_{f_1}, \ldots, \mathcal{D}_{f_m}]$, where each file-level diff $\mathcal{D}_{f_\ell}$ consists of a sequence of hunks $h_{f_\ell,*}$.

\paragraph{Divide the diffs.}
To synthesize the specification, we first conduct batch-level multi-label hunk matching (see Algorithm~\ref{alg:divide}).
Specifically, the agent $\pi_\theta^{\mathrm{map}}$ receives the task-structured release note $\mathcal{R}$ and a batch of file-level diffs $\mathcal{B}_j \subseteq \mathcal{D}$, and assigns each hunk to one or more task labels.
The label space is defined as $\mathcal{L} = \{1, \ldots, n\} \cup \{\texttt{doc}, \texttt{others}\}$, where labels $1,\ldots,n$ correspond to tasks $\mathcal{T}^{(1)}, \ldots, \mathcal{T}^{(n)}$, $\texttt{doc}$ denotes documentation-only changes, and $\texttt{others}$ denotes residual changes that cannot be aligned to any task.
Let $\mathcal{H}(\mathcal{B}_j)$ denote the hunks in batch $\mathcal{B}_j$. The batch-level matching result is
\begin{equation}
\Phi_j = \pi_\theta^{\mathrm{map}}(\mathcal{R}, \mathcal{B}_j), \qquad
\Phi_j : \mathcal{H}(\mathcal{B}_j) \rightarrow 2^{\mathcal{L}}.
\label{eq:map}
\end{equation}

After processing all batches, we merge the batch-level mappings into a global assignment $\Phi$ and reconstruct the matched hunk set $\mathcal{M}^{(k)}$ for task $\mathcal{T}^{(k)}$ by
\begin{equation}
    \mathcal{M}^{(k)} = \{\, h \in \mathcal{H}(\mathcal{D}) \;:\; k \in \Phi(h) \,\}.
\end{equation}

\paragraph{Synthesize the specifications.}
Given the matched hunk set $\mathcal{M}^{(k)}$, the agent $\pi_\theta^{\mathrm{syn}}$ then generates a specification $\mathcal{S}^{(k)}$ that describes the upgrade requirement for task $\mathcal{T}^{(k)}$ (see Algorithm~\ref{alg:conquer}):
\begin{equation}
    \mathcal{S}^{(k)} = \pi_\theta^{\mathrm{syn}}(\mathcal{T}^{(k)}, \mathcal{M}^{(k)}).
\end{equation}
For hunks assigned to $\texttt{others}$, the agent $\pi_\theta^{\mathrm{syn}}$ determines whether they collectively imply an additional upgrade task that is not explicitly covered by the release note.
To improve reliability and accommodate diverse levels of granularity, we adopt an oracle-like synthesis procedure inspired by \citet{zhuo2025bigcodebench} in which the agent is requested to generate the specification containing (1) problem statement, (2) grounded/conceptual expectations and constraints, (3) behaviors, and (4) acceptance criteria (see Prompt~\ref{prompt:synthesis_cont} for details).
We study their granularity effect in \S\ref{sec:res:granularity}.

\paragraph{Synthesis scale.}
In total, we use GPT-5.4~\cite{openai2026gpt54} (Codex~\cite{openai2025codex}) with \texttt{xhigh} reasoning effort as the agent to match the hunks and synthesize the specifications.
For the divide stage, we split file-level diffs into batches of at most 20 files, 150 hunks, and 3,000 diff lines.
Using Prompts~\ref{prompt:matching}, \ref{prompt:synthesis}, \ref{prompt:synthesis_cont}, and~\ref{prompt:complete} in Appendix~\ref{app:pro:synthesis}, we process 155 version transitions across 12 upgrade chains and synthesize 1,660 grounded upgrade tasks.
See Appendix~\ref{app:exa:synthesis} for examples of synthesis artifacts.

\paragraph{Stability check.}
Since the divide stage is the most susceptible to stochastic variation in our synthesis pipeline, we assess the stability on the largest diff (Pytest ver. 8.2.2 $\rightarrow$ 8.3.0), which contains 13k lines, 102 files, and 986 hunks.
Across 10 repeated hunk-matching runs, the assignments achieve 93.66\% exact match, 94.54\% Jaccard similarity, and 95.38\% core agreement, suggesting that the agent \emph{remains stable even under the most challenging transition}.
See Appendix~\ref{app:stability} for more details.

\subsection{Version-Upgrade Protocol}
\label{sec:met:version}
Each chain $\mathcal{C}$ is an ordered list of version-upgrade steps $[\mathcal{V}_1, \ldots, \mathcal{V}_N]$.
Starting from the initial codebase $c_{v_0}$ at version $v_0$ ($\hat{c}_{v_0}=c_{v_0}$), the agent $\pi_\theta$ sequentially performs each upgrade step conditioned on the synthesized specification $\mathcal{S}_i=\{\mathcal{S}_{i}^{(k)}\}_{k=1}^{K_i}$ (see Algorithm~\ref{alg:protocol}):
\begin{equation}
    \hat{c}_{v_i} = \pi_\theta(\mathcal{S}_{i}, \hat{c}_{v_{i-1}}), \qquad i = 1, \ldots, N.
\end{equation}

\paragraph{Build+Fix regularization.}
To avoid unfairly penalizing agents for non-behavioral execution incompatibilities, such as import, collection, or other setup-related failures, we design an additional regularization step.
This design is inspired by the multi-attempt setting (Pass Rate 2) used in Aider benchmark~\cite{aider2024aider}.
Specifically, we allow the agent to fix the codebase once to make it compatible with the execution setup when execution-level errors occur.

\section{Experimental Setup}
\label{sec:experimental}

\subsection{Environment and Configuration}
\label{sec:exp:environment}
\paragraph{Container environment and workspace.}
For each chain, we build a container from the starting-version environment and place the workspace under \texttt{/app}.
We then install the agent CLI inside the container, where it runs as a non-root user in \texttt{/app} and persists throughout the entire chain, rather than rebuilding a fresh environment at each step.
After each completed step, the codebase and specification are archived in a root-owned directory, keeping the active workspace clean and preventing the agent from being distracted by previous artifacts.

\paragraph{Anti-cheating controls.}
We leverage multiple anti-cheating controls to prevent shortcutting and information leakage.
At the agent level, we explicitly instruct them not to access or copy the released source code and enforce a shared set of tool-use restrictions for all agents, such as disabling web search and package installation commands (see Appendix~\ref{app:forbidden}).
At the sandbox level, we blacklist some code-hosting and package-registry domains, such as \texttt{githubusercontent.com} and \texttt{pypi.org}.

\subsection{Model and Agent Configuration}
\label{sec:exp:model}
We evaluate eight frontier models from four providers:
OpenAI (GPT-5.3-Codex~\cite{openai2026gpt53codex}, GPT-5.4~\cite{openai2026gpt54}, and GPT-5.5~\cite{openai2026gpt55}),
Anthropic (Claude-Sonnet-4.6~\cite{anthropic2026claudesonnet46}, Claude-Opus-4.6~\cite{anthropic2026claudeopus46}, and Claude-Opus-4.7~\cite{anthropic2026claudeopus47}),
MiniMax (MiniMax-M2.7-HS, the high-speed variant~\cite{minimax2026minimaxm27}), and
Z.AI (GLM-5.1~\cite{zai2026glm51}) on three agent CLIs: Codex~\cite{openai2025codex}, Claude Code~\cite{anthropic2025claudecode}, and OpenCode~\cite{opencode2025opencode}.
We use the \texttt{xhigh} reasoning effort for the two strongest models, Claude-Opus-4.7 and GPT-5.5, while all other models use \texttt{high} reasoning effort when available.
We conduct all experiments between April 17 and May 5, 2026, using the following agent CLI versions: Claude Code (2.1.113 - 2.1.126), Codex CLI (0.121.0 - 0.128.0), and OpenCode (1.14.18 - 1.14.32).

\paragraph{Agent invocation.}
Following Terminal-Bench~\cite{merrill2026terminalbench}, we invoke all agents in headless mode through a unified harness, which is useful for benchmarking raw model capabilities~\cite{orlanski2026slopcodebench}.
For each step, we place the specification and the agent-produced codebase under \texttt{/app}, then start a new agent session to avoid distractions from history compactions and better reflect realistic long-horizon maintenance, where developers return to a project after time away rather than continuing from an unbounded interactive context~\cite{orlanski2026slopcodebench}.
To keep prompts concise, we use the prompts that specify only the paths to the documents and the source tree, requiring the agent to read the relevant information directly from the workspace rather than embedding it in the prompt (see Prompts~\ref{prompt:build} and \ref{prompt:fix} in Appendix~\ref{app:pro:evaluation} and examples in Appendix~\ref{app:examples}).

\subsection{Evaluation}
\label{sec:exp:evaluation}
\paragraph{Evaluation configuration and protocol.}
We use the same evaluation protocol for cross-version validation in \S\ref{sec:met:benchmark}, hidden check, and post hoc evaluation, differing only in whether the codebase is extracted from the agent container or reconstructed by replaying code changes from the starting version.
Both are conducted in an isolated container built from the target package version, preventing contamination or leakage from the agent workspace.
To improve the robustness of pytest to the partial breakage typical of mid-upgrade code, we design collection and import plugins to ensure test coverage.
Concretely, unavailable APIs imported by test files are stubbed and logged, and phase-level reports are continuously written to disk so that collection failures and abnormal session termination still produce usable reports.
Additionally, we override the target version's pytest configuration to remove incompatible options, preventing shortcuts via test configuration edits for evaluation.

\paragraph{Evaluation metrics.}
For each upgrade step $\mathcal{V}_i$, we evaluate all tests in $\mathcal{Q}_{v_i}$ on the agent's previous and current codebases, $\hat{c}_{v_{i-1}}$ and $\hat{c}_{v_i}$.
Using the pre-defined upgrade-related set $\mathcal{U}_i$ in Eq.~\ref{eq:upgrade-related}, each test is classified by whether $q \in \mathcal{U}_i$ and by its status transition from the previous to the current codebase.
Table~\ref{tab:classification} summarizes the resulting categories.
We compute three metrics for each chain and report their averages across chains in \S\ref{sec:res:overall}:
\begin{itemize}[leftmargin=*]
\item \textbf{Resolving}: measures the agent's ability to implement the required upgrade behavior;
\item \textbf{Precision}: measures maintenance safety by penalizing regressions on previously passing; and
\item \textbf{F1-score}: balances upgrade resolving and behavior preservation.
\end{itemize}

Since upgrade steps vary in scope, we micro-average test outcomes across steps before computing each chain-level metric:
\begin{equation*}
    \mathrm{Resolving} = \frac{\sum_{i=1}^{n} \mathrm{TP}_i}{\sum_{i=1}^{n} (\mathrm{TP}_i + \mathrm{FN}_i)},
    \qquad
    \mathrm{Precision} = \frac{\sum_{i=1}^{n} \mathrm{TP}_i}{\sum_{i=1}^{n} (\mathrm{TP}_i + \mathrm{FP}_i)},
\end{equation*}
\begin{equation*}
    \mathrm{F1} = \frac{2\times\sum_{i=1}^{n}\mathrm{TP}_i}{\sum_{i=1}^{n}(2\mathrm{TP}_i+\mathrm{FP}_i+\mathrm{FN}_i)}.
\end{equation*}

\begin{table}[t]
\centering
\small
\caption{
Test classification at upgrade step $\mathcal{V}_i$ over all tests in $\mathcal{Q}_{v_i}$.
Upgrade-related tests follow $\mathcal{U}_i$ in Eq.~\ref{eq:upgrade-related}.
Here, $\mathrm{fail}$ includes assertion failures and execution errors, and ``$\cdot$'' denotes a wildcard.
}
\label{tab:classification}
\begin{tabular}{l c c c}
\toprule
\textbf{Category} & \textbf{Upgrade-related} & \textbf{Results $(\hat{c}_{v_{i-1}}, \hat{c}_{v_i})$} & \textbf{Symbol} \\
\midrule
resolved      & Yes & $(\cdot,\; \mathrm{pass})$              & $\mathrm{TP}_i$ \\
unresolved    & Yes & $(\cdot,\; \mathrm{fail})$              & $\mathrm{FN}_i$ \\
\midrule
preserved     & No  & $(\mathrm{pass},\; \mathrm{pass})$      & $\mathrm{TN}_i$ \\
regressed     & No  & $(\mathrm{pass},\; \mathrm{fail})$      & $\mathrm{FP}_i$ \\
\midrule
recovered     & No  & $(\mathrm{fail},\; \mathrm{pass})$      & $\mathrm{REC}_i$ \\
unrecovered   & No  & $(\mathrm{fail},\; \mathrm{fail})$      & $\overline{\mathrm{REC}}_i$ \\
\bottomrule
\end{tabular}
\end{table}

\section{Results}
\label{sec:results}

\begin{table}[t]
\centering
\caption{Average performance across 12 chains. 
\claudecode, \codex, and \opencode\ denote Claude Code, Codex, and OpenCode; 
\anthropic, \openai, \zai, and \minimax\ denote Anthropic, OpenAI, Z.AI, and MiniMax.}
\label{tab:avg_results}
\setlength{\tabcolsep}{6pt}
\renewcommand{\arraystretch}{1.25}
\setlength{\extrarowheight}{2pt}
\resizebox{1.0\linewidth}{!}{
\begin{tabular}{c c l | cc cc | cc}
\toprule
\multirow{2}{*}{\textbf{Agent}}
& \multirow{2}{*}{\textbf{Provider}}
& \multirow{2}{*}{\textbf{Model}}
& \multicolumn{2}{c}{\textbf{Resolving}}
& \multicolumn{2}{c|}{\textbf{Precision}}
& \multirow{2}{*}{\textbf{F1-score}}
& \multirow{2}{*}{\textbf{Final Passing}} \\
\cmidrule(lr){4-5} \cmidrule(lr){6-7}
& & & Build & Build + Fix & Build & Build + Fix & & \\
\midrule
\claudecode & \anthropic & Claude-Opus-4.7   & $\mathbf{58.3\%}$ & $\mathbf{60.8\%}$ & $67.4\%$ & $\mathbf{80.6\%}$ & $\mathbf{68.5\%}$ & $\underline{91.9\%}$ \\
\claudecode & \anthropic & Claude-Opus-4.6   & $38.8\%$ & $44.3\%$ & $51.3\%$ & $64.9\%$ & $50.8\%$ & $85.8\%$ \\
\claudecode & \anthropic & Claude-Sonnet-4.6 & $34.4\%$ & $39.8\%$ & $53.4\%$ & $66.5\%$ & $45.9\%$ & $84.5\%$ \\
\codex      & \openai    & GPT-5.5           & $\underline{52.2\%}$ & $\underline{57.5\%}$ & $\mathbf{72.0\%}$ & $\underline{80.1\%}$ & $\underline{64.8\%}$ & $\mathbf{92.2\%}$ \\
\codex      & \openai    & GPT-5.4           & $45.3\%$ & $47.5\%$ & $59.2\%$ & $72.2\%$ & $54.9\%$ & $90.4\%$ \\
\codex      & \openai    & GPT-5.3-Codex     & $47.1\%$ & $51.8\%$ & $\underline{70.5\%}$ & $76.3\%$ & $59.2\%$ & $90.9\%$ \\
\opencode   & \openai    & GPT-5.4           & $39.3\%$ & $43.4\%$ & $54.7\%$ & $63.8\%$ & $46.5\%$ & $89.7\%$ \\
\opencode   & \zai       & GLM-5.1           & $27.5\%$ & $38.1\%$ & $31.8\%$ & $49.9\%$ & $40.1\%$ & $84.1\%$ \\
\opencode   & \minimax   & MiniMax-M2.7-HS   & $14.5\%$ & $20.2\%$ & $25.8\%$ & $34.2\%$ & $21.2\%$ & $70.3\%$ \\
\midrule
\multicolumn{3}{c|}{\textbf{Average}}
& $39.7\%$ & $44.8\%$ & $54.0\%$ & $65.4\%$ & $50.2\%$ & $86.6\%$ \\
\bottomrule
\end{tabular}
}
\end{table}

\subsection{Overall Performance on \methodname}
\label{sec:res:overall}
Table~\ref{tab:avg_results} presents the overall results across 9 agent-model configurations under two regimes.
Across all agents and chains, the average Build+Fix resolving rate is only $44.8\%$, with $65.4\%$ precision and $50.2\%$ F1-score.
The higher precision suggests that successful changes are relatively targeted and non-regressive, while the resolving rate indicates that agents still leave more than half of the upgrade-related test behaviors unresolved, highlighting the challenge of continuous version-upgrade maintenance for current agents.
Nevertheless, Claude-Opus-4.7 (Claude Code) achieves the best performance, with $60.8\%$ resolving, $80.6\%$ precision, and $68.5\%$ F1-score, followed by GPT-5.5 (Codex).
We also report Final Passing as an auxiliary measure of final codebase health over the chain, which broadly follows the F1-score trend.
The substantially stronger performance of frontier agents shows that \methodname is a feasible and discriminative benchmark.

We also observe three clear patterns:
(1) frontier closed-source models outperform the other models in their respective families, and closed-source models generally outperform GLM-5.1 and MiniMax-M2.7-HS;
(2) GPT-5.3-Codex slightly outperforms GPT-5.4 under the Codex CLI, suggesting that Codex-native agentic variants can remain competitive with newer general-purpose models in continuous maintenance; and
(3) GPT-5.4 performs better with Codex than OpenCode, suggesting better alignment between the native agent CLI and its underlying model.
These patterns suggest that \methodname can distinguish differences in model capability, coding specialization, and agent CLIs.

Across agents, Build+Fix improves average precision from $54.0\%$ to $65.4\%$, while average resolving increases more modestly from $39.7\%$ to $44.8\%$, indicating that the additional fixing mainly addresses brittle failures and regressions rather than directly solving upgrade requirements.
Since reasonable implementations in black-box evaluation may still trigger broad failures due to setup incompatibilities (\eg, import errors or dependency issues), our Build+Fix regime mitigates this over-penalization by allowing a single correction based on error feedback, such as collection or import errors.
However, this does not make the benchmark easy, as the average resolving rate remains at only $44.8\%$, and weaker models like MiniMax-M2.7-HS still fall far below the initial Build performance of frontier models.
Notably, GPT-5.5 and GPT-5.3-Codex already achieve high Build precision and remain stable after Build+Fix, suggesting that they tend to produce safer first-pass patches, whereas Claude-Opus-4.7 achieves a Build resolving rate higher than the Build+Fix resolving rate of all other agents, showing stronger first-pass upgrade coverage.
See Appendix~\ref{app:trajectories} for Build and Build+Fix trajectories.

\begin{AIbox}{\linewidth}{Key Finding: \textbf{Overall Performance}}
{
\begin{itemize}[leftmargin=*]
    \item SWE-Chain is challenging and highly discriminative: agents achieve only $44.8\%$ resolving, $65.4\%$ precision, and $50.2\%$ F1-score on average under the Build+Fix regime.
    \item Claude-Opus-4.7 achieves the best performance in \methodname, followed by GPT-5.5.
    \item Build+Fix is necessary to prevent over-penalization due to incompatible setups, improving precision and leaving capability gaps intact.
\end{itemize}
}
\end{AIbox}

\subsection{Cross-Chain Difficulty}
\label{sec:res:cross}
\begin{figure}[t]
    \centering
    \includegraphics[width=1.0\linewidth]{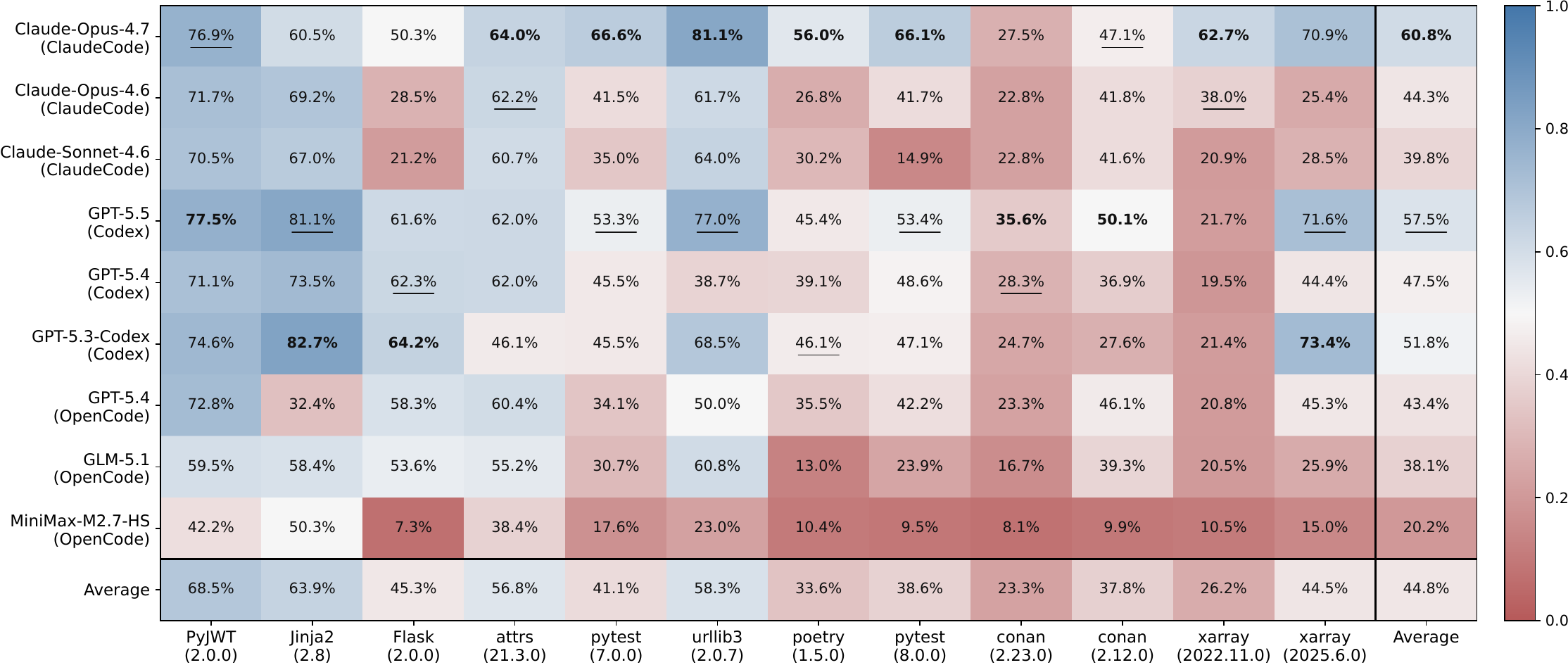}
    \caption{Build+Fix resolving rate by agent and chain, order left-to-right by increasing difficulty.}
    \label{fig:heatmap_resolving}
\end{figure}

Figure~\ref{fig:heatmap_resolving} illustrates the heatmap of Build+Fix resolving rate by chain and shows that performance varies sharply across chains.
Even on easier chains (\eg, \texttt{PyJWT} and \texttt{Jinja2}), agents resolve only below 70\% of upgrade-specific behaviors on average, while harder chains (\eg, \texttt{conan} and \texttt{xarray}) can fall below 30\%.
The codebase scale shown in Table~\ref{tab:chain_overview} partially explains this trend, but it is insufficient.
For example, \texttt{poetry} is smaller than \texttt{pytest}, but agents achieve lower resolving rates on \texttt{poetry} because its upgrades have a larger average code-diff scale.
The same pattern appears between \texttt{Flask} and \texttt{Jinja2}, suggesting that chain difficulty depends jointly on package size, upgrade scale, and density.
We provide the detailed difficulty ranking in Appendix~\ref{app:scale}.
Furthermore, no agent dominates across all chains: Claude-Opus-4.7 is stronger on chains like \texttt{xarray} (2022.11.0) and \texttt{pytest}, whereas GPT-based agents lead on chains such as \texttt{Flask}, \texttt{Jinja2}, and \texttt{conan}, indicating model-specific strengths across package structures.
Moreover, open-weight models remain competitive on easier chains but degrade sharply on harder chains, supporting that \methodname is discriminative across different maintenance scenarios.
We also provide the corresponding F1 and precision heatmaps in Appendix~\ref{app:heatmap}, and the per-step resolving trajectories in Appendix~\ref{app:perstep}.

\begin{AIbox}{\linewidth}{Key Finding: \textbf{Difficulty Is Chain-Specific}}
{
\methodname captures diverse difficulty across chains: the chain-level average resolving rate ranges from 23.3\% to 68.5\%, which discriminates agents across different upgrade scenarios.
}
\end{AIbox}

\subsection{Effect of Specification Granularity}
\label{sec:res:granularity}
To study how the specification granularity affects agent performance, we construct five specification variants on the pytest (ver. 8.0) chain, with examples shown in Appendix~\ref{app:exa:granularity}:
\begin{itemize}[leftmargin=*]
    \item \textbf{L1}: Release notes with GitHub issues and PRs, simulating SWE-EVO setting~\cite{thai2026sweevo}.
    \item \textbf{L2}: Problem statements only.
    \item \textbf{L3} (default setting in \S\ref{sec:res:overall}): Problem statements with conceptual expectations and constraints.
    \item \textbf{L4}: Problem statements with grounded (concrete API/behavior details) expectations and constraints.
    \item \textbf{L5}: Problem statements, grounded expectations and constraints, behaviors, and acceptance criteria.
\end{itemize}

As shown in Table~\ref{tab:spec_granularity}, the largest gap appears from L2 to L3, where expectations and constraints improve precision much more than resolving rate for both agents, as they clarify the intended behavior, implementation boundary, and preservation constraints, making the specification more informative and feasible for agents to conduct the version transition.
While L1 uses raw changelog entries with linked GitHub issues and PRs, which contain the issues-driven requirements and useful evidence, it is noisy, unevenly structured, and may expose implementation details.
Meanwhile, higher-granularity specifications provide more grounded information and generally improve performance, but they are closer to oracle-style guidance, making them better suited for assessing agents' implementation capability than realistic upgrade workflows.
However, they remain useful beyond upper-bound probes, especially for reliability evaluations where the objective is to sustain agent progress over many steps without excessive early collapse.

\begin{table}[t]
\centering
\caption{Effect of specification granularity on the pytest 8.0 chain under the Build+Fix regime.}
\label{tab:spec_granularity}
\small
\setlength{\tabcolsep}{6pt}
\renewcommand{\arraystretch}{1.15}
\begin{tabular}{l l c c c c c}
\toprule
\textbf{Model (Agent)} & \textbf{Metric} & \textbf{L1} & \textbf{L2} & \textbf{L3} & \textbf{L4} & \textbf{L5} \\
\midrule
\multirow{3}{*}{Claude-Opus-4.7 \claudecode} 
& Resolving  & $44.5\%$ & $62.6\%$ & $66.1\%$ & $66.7\%$ & $73.9\%$ \\
& Precision  & $8.9\%$  & $7.8\%$  & $88.1\%$ & $89.2\%$ & $93.5\%$ \\
& F1         & $14.8\%$ & $13.9\%$ & $75.5\%$ & $76.3\%$ & $82.5\%$ \\
\cmidrule(lr){1-7}
\multirow{3}{*}{GPT-5.5 \codex} 
& Resolving  & $47.4\%$ & $44.5\%$ & $53.4\%$ & $68.1\%$ & $67.8\%$ \\
& Precision  & $9.9\%$  & $9.5\%$  & $92.1\%$ & $93.3\%$ & $89.7\%$ \\
& F1         & $16.3\%$ & $15.7\%$ & $67.6\%$ & $78.7\%$ & $77.3\%$ \\
\bottomrule
\end{tabular}
\end{table}

\begin{AIbox}{\linewidth}{Key Finding: \textbf{Granularity Control}}
{
Raw issue/PR artifacts are poor specifications for release-level upgrades, while expectations and constraints are essential to avoid regressions.
Our default L3 setting balances guidance and leakage control without relying on the more oracle-like information in L4 and L5.
}
\end{AIbox}

\subsection{Efficiency and Resource Usage}
\label{sec:res:efficiency}

Table~\ref{tab:efficiency_usage} reports the efficiency, cost, token usage, and tool-call activity of each agent averaged per chain.
The strongest configurations with \texttt{xhigh} reasoning effort are also among the most expensive: Claude-Opus-4.7 and GPT-5.5 lead in performance but cost $\$150.39$ and $\$131.34$ per chain on average.
However, higher resource usage does not inevitably imply better results.
For example, Claude-Sonnet-4.6 has the longest runtime and high token usage, and MiniMax-M2.7-HS uses more tokens and tool calls than OpenCode GPT-5.4, but both perform substantially worse than top models.
Furthermore, tool-call success rates are consistently high across models, the performance gaps are unlikely to be driven by tool failures.
Overall, \methodname measures maintenance capability rather than merely interaction volume or execution reliability.

\begin{table*}[t]
\centering
\caption{Average efficiency, cost, token usage and tool-use statistics per chain. Runtime measures the agent execution time for the chain (excluding hidden-test execution). Maximum values are in \textbf{\textcolor{red}{red\ bold}}; minimum values are in \textcolor{blue!50!black}{\underline{blue underlined}}.}
\label{tab:efficiency_usage}
\small
\setlength{\tabcolsep}{6pt}
\renewcommand{\arraystretch}{1.15}
\resizebox{0.9\linewidth}{!}{
\begin{tabular}{c c l | c c c c | c c}
\toprule
\multirow{2}{*}{\textbf{Agent}}
& \multirow{2}{*}{\textbf{Provider}}
& \multirow{2}{*}{\textbf{Model}}
& \multicolumn{2}{c}{\textbf{Runtime (h)}} 
& \multirow{2}{*}{\textbf{Cost}}
& \multirow{2}{*}{\textbf{Tokens}}
& \multicolumn{2}{c}{\textbf{Tools}}  \\
\cmidrule(lr){4-5} \cmidrule(lr){8-9}
& & & Build & Build+Fix & & & Calls & Success \\
\midrule
\claudecode & \anthropic & Claude-Opus-4.7
& $3.13$ & $3.43$ & $\mathbf{\textcolor{red}{\$150.39}}$ & $\mathbf{\textcolor{red}{350.7\text{M}}}$ & $1,748$ & $98.3\%$ \\
\claudecode & \anthropic & Claude-Opus-4.6
& $2.68$ & $3.25$ & $\$71.78$ & $179.4\text{M}$ & $\mathbf{\textcolor{red}{2,008}}$ & $98.4\%$ \\
\claudecode & \anthropic & Claude-Sonnet-4.6
& $\mathbf{\textcolor{red}{4.13}}$ & $\mathbf{\textcolor{red}{4.84}}$ & $\$58.68$ & $203.9\text{M}$ & $1,866$ & $\mathbf{\textcolor{red}{98.6\%}}$ \\
\codex & \openai & GPT-5.5
& $3.23$ & $3.48$ & $\$131.34$ & $184.7\text{M}$ & $1,736$ & $95.6\%$ \\
\codex & \openai & GPT-5.4
& $\underline{\textcolor{blue!50!black}{2.12}}$ & $\underline{\textcolor{blue!50!black}{2.28}}$ & $\$39.84$ & $100.9\text{M}$ & $1,165$ & $94.6\%$ \\
\codex & \openai & GPT-5.3-Codex
& $2.58$ & $2.77$ & $\$41.59$ & $155.2\text{M}$ & $1,318$ & $\underline{\textcolor{blue!50!black}{93.8\%}}$ \\
\opencode & \openai & GPT-5.4
& $2.17$ & $2.44$ & $\$22.98$ & $\underline{\textcolor{blue!50!black}{61.1\text{M}}}$ & $1,098$ & $95.4\%$ \\
\opencode & \zai & GLM-5.1
& $4.09$ & $4.44$ & $\$20.09$ & $67.5\text{M}$ & $\underline{\textcolor{blue!50!black}{1,057}}$ & $97.4\%$ \\
\opencode & \minimax & MiniMax-M2.7-HS
& $2.19$ & $2.68$ & $\underline{\textcolor{blue!50!black}{\$9.51}}$ & $115.0\text{M}$ & $1,432$ & $95.4\%$ \\
\bottomrule
\end{tabular}
}
\end{table*}

\section{Related Works}

\paragraph{AI4SE benchmarks.}
The evaluation of coding agents has advanced rapidly through a series of increasingly challenging benchmarks.
EvalPlus~\cite{liu2023evalplus}, BigCodeBench~\cite{zhuo2025bigcodebench}, and LiveCodeBench~\cite{jain2025livecodebench} evaluate functional correctness, library-level API usage, and contamination-free evaluation across diverse programming tasks.
SWE-Bench~\cite{jimenez2024swebench} pioneers agent evaluation on real-world GitHub issues, requiring models to localize faults and implement fixes across open-source repositories.
Its variants~\cite{openai2024swebenchverified, yang2024swebenchmultimodal, deng2025swebenchpro} further improve evaluation reliability, extend to visual artifacts, and increase task difficulty.
$\tau$-Bench~\cite{yao2025taubench} evaluates agents on tool-use and decision-making in multi-turn interactive environments, where agents must complete user-specified goals by invoking APIs and reasoning over structured databases under realistic constraints.
Terminal-Bench~\cite{merrill2026terminalbench} pushes further into agentic territory by assessing agents on long-horizon command-line tasks inside persistent shell sessions, exposing limitations in environment state tracking, tool chaining, and error recovery across interaction trajectories. 

\paragraph{Long-horizon maintenance benchmarks.}
Recent benchmarks evaluate coding agents beyond isolated tasks.
NL2RepoBench~\cite{ding2026nl2repobench}, MaintainCoder~\cite{wang2025maintaincoder}, VibeCodeBench~\cite{tran2026vibecodebench}, and CodeFlowBench~\cite{wang2026codeflowbench} study long-horizon generation, maintainability under changing requirements, and multi-step coding workflows, but they primarily focus on generation-oriented or requirement-driven development rather than continuous evolution.
SWE-EVO~\cite{thai2026sweevo} approaches repository evolution by evaluating agents on real software upgrades, but its tasks are still end-to-end, and their specifications remain coarse-grained when derived directly from GitHub issues and pull requests.
SlopCodeBench~\cite{orlanski2026slopcodebench} studies degradation when agents iteratively extend their own generated code, which is closer to repository construction like NL2RepoBench than to maintenance of mature human-developed codebases.
SWE-CI~\cite{chen2026sweci} evaluates maintenance using CI traces, and EvoClaw~\cite{deng2026evoclaw} reconstructs milestone-level evolution from commit histories.
While these works model important aspects of software engineering lifecycles, their task boundaries do not align with official package releases, where maintainers bundle, document, test, and ship version-level changes constantly.
Unlike prior benchmarks, \methodname segments evolution at maintainer-defined release nodes and uses fine-grained synthetic specifications to address the coarse granularity of release-level upgrades.

\section{Conclusion}
We introduce \methodname, a benchmark for chained release-level package upgrades, comprising 12 upgrade chains across diverse domains, scales, and difficulty levels to evaluate coding agents' codebase maintenance capabilities.
We design \synthesis that aligns maintainer release notes with gold code diffs to produce grounded, reliable, informative, and feasible upgrade specifications at scale.
Across 9 frontier configurations, agents achieve only $44.8\%$ resolving and $50.2\%$ F1 on average under Build+Fix.
These results show that current agents still struggle to carry their own changes forward across chained releases without breaking existing functionality, highlighting the need for agents that can maintain coherent, reliable, and evolvable codebases across long chains.

\clearpage

\bibliography{reference}
\bibliographystyle{plainnat}

\clearpage
\appendix

\tableofcontents
\addtocontents{toc}{\protect\setcounter{tocdepth}{2}}
\clearpage

\lstset{
    basicstyle=\ttfamily\scriptsize,
    columns=fullflexible,
    frame=single,
    breaklines=true,
    breakatwhitespace=true,
    postbreak=\mbox{$\hookrightarrow$\space},
    rulecolor=\color{black},
    showstringspaces=false,
    keepspaces=true,
    aboveskip=0.5em,
    belowskip=0.5em,
    xleftmargin=0.5em, xrightmargin=0.5em,
    numbers=none,
    keywordstyle=\color{black},
    commentstyle=\color{pythoncomment},
    stringstyle=\color{pythonstring}
}

\section{Notation}

\begin{table*}[ht]
\centering
\caption{Notation used in \methodname.}
\label{tab:notation}
\small
\setlength{\tabcolsep}{8pt}
\renewcommand{\arraystretch}{1.15}
\begin{tabular}{c p{0.72\linewidth}}
\toprule
\textbf{Symbol}      & \textbf{Meaning} \\
\midrule

\multicolumn{2}{c}{\textbf{Version and Codebase Notation}} \\
\midrule
$i$                  & Index of a version-upgrade step in an upgrade chain. \\
$v_i$                & Version identifier at step $i$. \\
$\mathcal{V}_i$      & Version-upgrade step from version $v_{i-1}$ to $v_i$. \\
$c_{v_i}$            & Gold codebase of version $v_i$. \\
$\hat{c}_{v_i}$      & Agent-produced codebase after upgrading to version $v_i$. \\
$\hat{c}_{v_0}$      & Initial agent codebase, initialized as the grounded codebase $c_{v_0}$. \\
$\mathcal{C}$        & An upgrade chain consisting of sequential version-upgrade steps $[\mathcal{V}_1,\ldots,\mathcal{V}_N]$. \\
$\mathcal{Q}_{v_i}$  & Test suite of version $v_i$. \\
$q$                  & A test case in a version test suite. \\
$r(q,c)$             & Execution result of test case $q$ on codebase $c$, taking values in $\{\mathrm{pass}, \mathrm{fail}, \mathrm{error}\}$. \\

\midrule
\multicolumn{2}{c}{\textbf{Release Note and Diff Notation}} \\
\midrule
$\mathcal{R}$              & Release note or changelog represented as a list of tasks $[\mathcal{T}^{(1)}, \ldots, \mathcal{T}^{(n)}]$. \\
$\mathcal{T}^{(k)}$        & The $k$-th task extracted from the release note. \\
$\mathcal{D}$              & Diff context between two consecutive versions, consisting of code and test diffs. \\
$\mathcal{D}_{f_\ell}$     & File-level diff for file $f_\ell$. \\
$h_{f_\ell,*}$             & Sequence of hunks contained in file-level diff $\mathcal{D}_{f_\ell}$. \\
$\mathcal{H}(\mathcal{D})$ & Set of all hunks in the diff context $\mathcal{D}$. \\

\midrule
\multicolumn{2}{c}{\textbf{Specification Synthesis Notation}} \\
\midrule
$\mathcal{B}_j$              & A batch of file-level diffs used for batch-level hunk matching. \\
$\mathcal{H}(\mathcal{B}_j)$ & Set of hunks contained in diff batch $\mathcal{B}_j$. \\
$\mathcal{L}$                & Label space for hunk matching, defined as $\{1,\ldots,n\}\cup\{\texttt{doc},\texttt{others}\}$. \\
$\Phi_j$                     & Batch-level mapping from hunks in $\mathcal{H}(\mathcal{B}_j)$ to one or more labels in $\mathcal{L}$. \\
$\Phi$                       & Global hunk-to-label assignment obtained by merging all batch-level mappings. \\
$\mathcal{M}^{(k)}$          & Matched hunk set for release-note task $\mathcal{T}^{(k)}$. \\
$\mathcal{S}^{(k)}$          & Synthesized specification for task $\mathcal{T}^{(k)}$. \\
$\pi_\theta^{\mathrm{map}}$  & Agent used for batch-level multi-label hunk matching. \\
$\pi_\theta^{\mathrm{syn}}$  & Agent used for specification synthesis. \\

\midrule
\multicolumn{2}{c}{\textbf{Version-Upgrade Evaluation Notation}} \\
\midrule
$\mathcal{U}_i$       & Upgrade-related test set for version-upgrade step $\mathcal{V}_i$. \\
$K_i$                 & Number of synthesized task specifications for version-upgrade step $\mathcal{V}_i$. \\
$\mathcal{S}_i^{(k)}$ & Synthesized specification for the $k$-th task at step $\mathcal{V}_i$. \\
$\mathcal{S}_i$       & Set of synthesized specifications for step $\mathcal{V}_i$, defined as $\{\mathcal{S}_{i}^{(k)}\}_{k=1}^{K_i}$. \\
$\pi_\theta$          & Evaluated coding agent that performs sequential version upgrades. \\

\midrule
\multicolumn{2}{c}{\textbf{Evaluation Outcome Categories}} \\
\midrule
$\mathrm{TP}_i$                  & Resolved upgrade-related tests at step $\mathcal{V}_i$ (passing on $\hat{c}_{v_i}$). \\
$\mathrm{FN}_i$                  & Unresolved upgrade-related tests at step $\mathcal{V}_i$ (failing on $\hat{c}_{v_i}$). \\
$\mathrm{TN}_i$                  & Preserved non-upgrade tests at step $\mathcal{V}_i$ (passing on both $\hat{c}_{v_{i-1}}$ and $\hat{c}_{v_i}$). \\
$\mathrm{FP}_i$                  & Regressed non-upgrade tests at step $\mathcal{V}_i$ (passing on $\hat{c}_{v_{i-1}}$, failing on $\hat{c}_{v_i}$). \\
$\mathrm{REC}_i$                 & Recovered non-upgrade tests at step $\mathcal{V}_i$ (failing on $\hat{c}_{v_{i-1}}$, passing on $\hat{c}_{v_i}$). \\
$\overline{\mathrm{REC}}_i$      & Unrecovered non-upgrade tests at step $\mathcal{V}_i$ (failing on both $\hat{c}_{v_{i-1}}$ and $\hat{c}_{v_i}$). \\
\bottomrule
\end{tabular}
\end{table*}

\clearpage
\section{Benchmark Scale and Difficulty Ranking}
\label{app:scale}

\begin{table*}[ht]
\centering
\caption{Source-tree size of the starting and ending package versions for each chain.
The ``no tests'' columns drop the package test folder, with sub-columns ``Python'' (\texttt{.py} only)
and ``All'' (all file types); the rightmost LOC$_{\text{start}}$ and LOC$_{\text{end}}$ count every file with no exclusions.}
\label{tab:chain_loc}
\resizebox{1.0\textwidth}{!}{
\renewcommand{\arraystretch}{1.5}
\begin{tabular}{llll|cccc|cc}
\toprule
\multirow{2}{*}{\textbf{Package}} & \multirow{2}{*}{\textbf{Repo}} & \multirow{2}{*}{\textbf{License}} & \multirow{2}{*}{\textbf{Chain}} & \multicolumn{2}{c}{\textbf{LOC$_{\text{start}}^{\text{no tests}}$}} & \multicolumn{2}{c|}{\textbf{LOC$_{\text{end}}^{\text{no tests}}$}} & \multirow{2}{*}{\textbf{LOC$_{\text{start}}$}} & \multirow{2}{*}{\textbf{LOC$_{\text{end}}$}} \\
\cmidrule(lr){5-6} \cmidrule(lr){7-8}
 & & & & Python & All & Python & All & & \\
\midrule
\href{https://pypi.org/project/attrs/}{attrs} & \href{https://github.com/python-attrs/attrs}{python-attrs/attrs} & MIT & {21.3.0 $\rightarrow$ 26.1.0} & $5.8\text{k}$ & $13\text{k}$ & $7.4\text{k}$ & $23\text{k}$ & $25\text{k}$ & $38\text{k}$ \\
\multirow[c]{2}{*}{\href{https://pypi.org/project/conan/}{conan}} & \multirow[c]{2}{*}{\href{https://github.com/conan-io/conan}{conan-io/conan}} & \multirow[c]{2}{*}{MIT} & {2.12.0 $\rightarrow$ 2.20.1} & $52\text{k}$ & $67\text{k}$ & $57\text{k}$ & $73\text{k}$ & $172\text{k}$ & $188\text{k}$ \\
 &  &  & {2.23.0 $\rightarrow$ 2.28.1} & $59\text{k}$ & $76\text{k}$ & $62\text{k}$ & $78\text{k}$ & $193\text{k}$ & $203\text{k}$ \\
\href{https://pypi.org/project/Flask/}{Flask} & \href{https://github.com/pallets/flask}{pallets/flask} & BSD-3-Clause & {2.0.0 $\rightarrow$ 2.3.3} & $9.0\text{k}$ & $29\text{k}$ & $9.6\text{k}$ & $29\text{k}$ & $36\text{k}$ & $37\text{k}$ \\
\href{https://pypi.org/project/Jinja2/}{Jinja2} & \href{https://github.com/pallets/jinja2}{pallets/jinja2} & BSD-3-Clause & {2.8 $\rightarrow$ 2.10.3} & $13\text{k}$ & $23\text{k}$ & $15\text{k}$ & $24\text{k}$ & $27\text{k}$ & $30\text{k}$ \\
\href{https://pypi.org/project/poetry/}{poetry} & \href{https://github.com/python-poetry/poetry}{python-poetry/poetry} & MIT & {1.5.0 $\rightarrow$ 1.8.5} & $21\text{k}$ & $21\text{k}$ & $22\text{k}$ & $23\text{k}$ & $81\text{k}$ & $87\text{k}$ \\
\href{https://pypi.org/project/PyJWT/}{PyJWT} & \href{https://github.com/jpadilla/pyjwt}{jpadilla/pyjwt} & MIT & {2.0.0 $\rightarrow$ 2.12.1} & $1.6\text{k}$ & $3.4\text{k}$ & $3.1\text{k}$ & $5.9\text{k}$ & $5.9\text{k}$ & $11\text{k}$ \\
\multirow[c]{2}{*}{\href{https://pypi.org/project/pytest/}{pytest}} & \multirow[c]{2}{*}{\href{https://github.com/pytest-dev/pytest}{pytest-dev/pytest}} & \multirow[c]{2}{*}{MIT} & {7.0.0 $\rightarrow$ 7.4.4} & $31\text{k}$ & $79\text{k}$ & $34\text{k}$ & $86\text{k}$ & $128\text{k}$ & $140\text{k}$ \\
 &  &  & {8.0.0 $\rightarrow$ 8.3.5} & $35\text{k}$ & $89\text{k}$ & $36\text{k}$ & $93\text{k}$ & $145\text{k}$ & $152\text{k}$ \\
\href{https://pypi.org/project/urllib3/}{urllib3} & \href{https://github.com/urllib3/urllib3}{urllib3/urllib3} & MIT & {2.0.7 $\rightarrow$ 2.6.3} & $13\text{k}$ & $17\text{k}$ & $13\text{k}$ & $21\text{k}$ & $32\text{k}$ & $40\text{k}$ \\
\multirow[c]{2}{*}{\href{https://pypi.org/project/xarray/}{xarray}} & \multirow[c]{2}{*}{\href{https://github.com/pydata/xarray}{pydata/xarray}} & \multirow[c]{2}{*}{Apache-2.0} & {2022.11.0 $\rightarrow$ 2023.7.0} & $76\text{k}$ & $119\text{k}$ & $81\text{k}$ & $131\text{k}$ & $179\text{k}$ & $192\text{k}$ \\
 &  &  & {2025.6.0 $\rightarrow$ 2026.2.0} & $104\text{k}$ & $152\text{k}$ & $108\text{k}$ & $159\text{k}$ & $230\text{k}$ & $244\text{k}$ \\
\bottomrule
\end{tabular}}
\end{table*}

Table~\ref{tab:chain_loc} reports the codebase size of every chain at its starting and ending package versions, with separate counts that drop the package test folder, which is removed in evaluation.

\paragraph{Difficulty ranking.}
We rank the 12 upgrade chains using two scale-based factors:
(1) the Python-only starting codebase size excluding tests, measured by $\mathrm{LOC}_{\mathrm{start}}^{\mathrm{py,\ no\ tests}}$; and
(2) the average code-change burden per upgrade step, measured by $\mathrm{LOCD}/\#\mathrm{Upgrades}$.
We min--max normalize both factors across the 12 chains and average them:
\begin{equation*}
\mathrm{Difficulty} = \frac{1}{2}\left(\mathrm{norm}\left(\mathrm{LOC}_{\mathrm{start}}^{\mathrm{py,\ no\ tests}}\right) + \mathrm{norm}\Big(\frac{\mathrm{LOCD}}{\#\mathrm{Upgrades}}\Big) \right).
\end{equation*}
Table~\ref{tab:chain_difficulty} shows the resulting difficulty ranking, where larger Python codebases and denser per-upgrade code changes generally lead to higher difficulty.

\begin{table}[ht]
\centering
\caption{Difficulty ranking in \methodname, which combines normalized starting codebase size and LOCD/\#Upgrades. Rank 1 is the hardest.}
\label{tab:chain_difficulty}
\resizebox{0.75\textwidth}{!}{
\renewcommand{\arraystretch}{1.4}
\begin{tabular}{cllccc}
\toprule
\textbf{Rank} & \textbf{Package} & \textbf{Chain} & \textbf{LOC$_{\text{start}}^{\text{py, no tests}}$} & \textbf{LOCD/\#Upgrades} & \textbf{Difficulty} \\
\midrule
$1$ & \href{https://pypi.org/project/xarray/}{xarray} & {2025.6.0 $\rightarrow$ 2026.2.0} & $104\text{k}$ & $1283.6$ & $0.989$ \\
$2$ & \href{https://pypi.org/project/xarray/}{xarray} & {2022.11.0 $\rightarrow$ 2023.7.0} & $76\text{k}$ & $1308.2$ & $0.866$ \\
$3$ & \href{https://pypi.org/project/conan/}{conan} & {2.12.0 $\rightarrow$ 2.20.1} & $52\text{k}$ & $868.8$ & $0.544$ \\
$4$ & \href{https://pypi.org/project/conan/}{conan} & {2.23.0 $\rightarrow$ 2.28.1} & $59\text{k}$ & $662.2$ & $0.484$ \\
$5$ & \href{https://pypi.org/project/pytest/}{pytest} & {8.0.0 $\rightarrow$ 8.3.5} & $35\text{k}$ & $860.7$ & $0.456$ \\
$6$ & \href{https://pypi.org/project/poetry/}{poetry} & {1.5.0 $\rightarrow$ 1.8.5} & $21\text{k}$ & $897.7$ & $0.404$ \\
$7$ & \href{https://pypi.org/project/urllib3/}{urllib3} & {2.0.7 $\rightarrow$ 2.6.3} & $13\text{k}$ & $678.4$ & $0.263$ \\
$8$ & \href{https://pypi.org/project/pytest/}{pytest} & {7.0.0 $\rightarrow$ 7.4.4} & $31\text{k}$ & $451.6$ & $0.246$ \\
$9$ & \href{https://pypi.org/project/attrs/}{attrs} & {21.3.0 $\rightarrow$ 26.1.0} & $5.8\text{k}$ & $602.6$ & $0.195$ \\
$10$ & \href{https://pypi.org/project/Flask/}{Flask} & {2.0.0 $\rightarrow$ 2.3.3} & $9.0\text{k}$ & $471.4$ & $0.150$ \\
$11$ & \href{https://pypi.org/project/Jinja2/}{Jinja2} & {2.8 $\rightarrow$ 2.10.3} & $13\text{k}$ & $394.2$ & $0.135$ \\
$12$ & \href{https://pypi.org/project/PyJWT/}{PyJWT} & {2.0.0 $\rightarrow$ 2.12.1} & $1.6\text{k}$ & $224.6$ & $0.000$ \\
\bottomrule
\end{tabular}}
\end{table}

\clearpage
\section{Stability of the Divide Step}
\label{app:stability}

As mentioned in \S\ref{sec:met:synthesis}, the divide step is the most susceptible to stochastic variation in \synthesis.
We run the matching agent $\pi_\theta^{\mathrm{map}}$ 10 times on the largest version transition, \texttt{pytest} 8.2.2 $\rightarrow$ 8.3.0, which contains 13k diff lines, 102 files, and 986 hunks, to model the most stressful situation.

By Eq.~\ref{eq:map}, each run $r$ produces a hunk-to-label mapping
\[
\Phi^{(r)}: \mathcal{H} \rightarrow 2^{\mathcal{L}},
\]
where $\mathcal{H}$ is the set of hunks and $\Phi^{(r)}(h)$ is the set of task labels assigned to hunk $h$ in run $r$.
We compare every pair of runs $(r,s)$ over the common hunk set $\mathcal{H}$ using three hunk-level similarity metrics.

\paragraph{Exact match.}
Exact match requires the assigned label sets to be identical:
\[
\mathrm{Exact}(r,s)
=
\frac{1}{|\mathcal{H}|}
\sum_{h\in\mathcal{H}}
\mathbf{1}\!\left[\Phi^{(r)}(h)=\Phi^{(s)}(h)\right].
\]

\paragraph{Jaccard similarity.}
Jaccard similarity measures partial overlap between assigned label sets:
\[
\mathrm{Jaccard}(r,s)
=
\frac{1}{|\mathcal{H}|}
\sum_{h\in\mathcal{H}}
\frac{
|\Phi^{(r)}(h)\cap \Phi^{(s)}(h)|
}{
|\Phi^{(r)}(h)\cup \Phi^{(s)}(h)|
},
\]
where two empty label sets are represented by a similarity of $1$.

\paragraph{Core agreement.}
Since providing additional hunks is less harmful than missing core task-relevant hunks, we define the core agreement as a superset-tolerant metric beyond Jaccard similarity.
Core agreement counts two runs as agreeing on a hunk if their assigned label sets overlap, or if both are empty.
\[
\mathrm{Core}(r,s)
=
\frac{1}{|\mathcal{H}|}
\sum_{h\in\mathcal{H}}
\mathbf{1}\!\left[
\left(\Phi^{(r)}(h)\cap \Phi^{(s)}(h)\neq \emptyset\right)
\lor
\left(\Phi^{(r)}(h)=\Phi^{(s)}(h)=\emptyset\right)
\right].
\]

We report the average over all $\binom{10}{2}=45$ run pairs.
The results are 93.66\% exact match, 94.54\% Jaccard similarity, and 95.38\% core agreement, showing that the divide step remains stable even in the most challenging situation.

\clearpage
\section{Forbidden Tools}
\label{app:forbidden}
As described in \S\ref{sec:exp:environment}, we restrict the tools and commands available to agents during task execution in Artifact~\ref{artifact:forbidden} to prevent external dependency installation and remote repository access.
Agents must complete each upgrade using only the files, packages, and tools already available in the local environment.

\begin{artifact}[ht]
\begin{AIBox}{Forbidden Tools List}
{\ttfamily\scriptsize
\textbf{Web access and remote fetching:}\\
- WebFetch\\
- WebSearch\\
- curl / wget\\
\\
\textbf{Git remote operations:}\\
- git fetch\\
- git pull\\
- git clone\\
- git push\\
- git ls-remote\\
- git remote\\
- git submodule\\
- git archive --remote\\
\\
\textbf{Remote shell and file transfer:}\\
- ssh\\
- scp\\
- sftp\\
- rsync\\
- nc / netcat\\
- telnet\\
- openssl s\_client\\
\\
\textbf{Package installation and dependency downloading:}\\
- pip install / pip download / pip wheel\\
- python -m pip install / download / wheel\\
- python3 -m pip install / download / wheel\\
- uv pip install / download / compile\\
- poetry add / poetry install\\
- npm / yarn / pnpm\\
- apt / apt-get\\
- conda install / mamba install / micromamba install\\
}
\end{AIBox}
\caption{Forbidden tools and commands during agent execution.}
\label{artifact:forbidden}
\end{artifact}

\clearpage
\section{Pseudocode for \methodname}
\label{app:pseudocode}
This appendix gives the pseudocode for the three procedures of \methodname:
the divide step (Algorithm~\ref{alg:divide}) and conquer step (Algorithm~\ref{alg:conquer}) of \synthesis in \S\ref{sec:met:synthesis}, and the version-upgrade evaluation protocol (Algorithm~\ref{alg:protocol}) in \S\ref{sec:met:version}.

\begin{algorithm}[ht]
\caption{\synthesis (Divide): Hunk-to-task matching.}
\label{alg:divide}
\begin{algorithmic}[1]
\Require Release note $\mathcal{R}$; file-level diffs $\mathcal{D}$; matching agent $\pi_\theta^{\mathrm{map}}$; retry budget $q$
\Ensure Per-task hunk lists, plus \texttt{doc} and \texttt{others} buckets
\State $\mathcal{D} \leftarrow \mathrm{SortByLineCount}(\mathcal{D})$ \Comment{descending}
\State $\{\mathcal{B}_j\}_{j=1}^{J} \leftarrow \mathrm{GreedyPack}(\mathcal{D};\, 20\ \text{files},\,150\ \text{hunks},\,3000\ \text{diff lines})$ \Comment{all limits must hold}
\State $\Phi \leftarrow \emptyset$
\For{each batch $\mathcal{B}_j$} \Comment{batch-level hunk matching}
    \State $H_j \leftarrow \mathrm{ExtractHunks}(\mathcal{B}_j)$
    \For{$t=1$ to $q$} \Comment{default: $q=3$}
        \State $\hat{\Phi}_j \leftarrow \pi_\theta^{\mathrm{map}}(\mathcal{R}, \mathcal{B}_j)$
        \State $(H_{\mathrm{miss}}, H_{\mathrm{spur}}) \leftarrow \mathrm{CheckCoverage}(\hat{\Phi}_j, H_j)$ \Comment{extract missing or invented hunks}
        \If{$H_{\mathrm{miss}}=\emptyset$ and $H_{\mathrm{spur}}=\emptyset$}
            \State $\Phi \leftarrow \Phi \cup \hat{\Phi}_j$
            \State \textbf{break}
        \EndIf
        \State $\mathcal{B}_j \leftarrow \mathrm{AppendFeedback}(\mathcal{B}_j, H_{\mathrm{miss}}, H_{\mathrm{spur}})$
    \EndFor
\EndFor
\State $\mathcal{M} \leftarrow \mathrm{GroupByLabel}(\Phi)$ \Comment{tasks, \texttt{doc}, and \texttt{others}}
\State \Return $\mathcal{M}$
\end{algorithmic}
\end{algorithm}

\begin{algorithm}[ht]
\caption{\synthesis (Conquer): Per-task specification synthesis.}
\label{alg:conquer}
\begin{algorithmic}[1]
\Require Release note $\mathcal{R}$; matched hunk buckets $\mathcal{M}$; synthesis agent $\pi_\theta^{\mathrm{syn}}$; retry budget $q$
\Ensure Structured task specifications $\mathcal{S}$
\State $\mathcal{S} \leftarrow \emptyset$
\For{each release-note task $\mathcal{T}^{(k)} \in \mathcal{R}$ \textbf{in parallel}} \Comment{task-level synthesis}
    \State $\mathcal{M}^{(k)} \leftarrow \mathrm{GetMatchedHunks}(\mathcal{M}, k)$
    \State $\mathcal{I}^{(k)} \leftarrow (\mathcal{T}^{(k)}, \mathcal{M}^{(k)})$
    \For{$t=1$ to $q$} \Comment{default: $q=3$}
        \State $\hat{\mathcal{S}}^{(k)} \leftarrow \pi_\theta^{\mathrm{syn}}(\mathcal{I}^{(k)})$
        \If{$\mathrm{ValidateSchema}(\hat{\mathcal{S}}^{(k)})$}
            \State $\mathcal{S} \leftarrow \mathcal{S} \cup \{\hat{\mathcal{S}}^{(k)}\}$
            \State \textbf{break}
        \EndIf
        \State $\mathcal{I}^{(k)} \leftarrow \mathrm{AppendSchemaFeedback}(\mathcal{I}^{(k)})$
    \EndFor
\EndFor
\If{$\mathcal{M}^{(\texttt{others})} \neq \emptyset$} \Comment{handle \texttt{others} labeled hunks}
\State $\mathcal{I}^{(\texttt{new})} \leftarrow (\mathcal{S}, \mathcal{M}^{(\texttt{others})})$
    \For{$t=1$ to $q$}
        \State $\hat{\mathcal{S}}^{(\texttt{new})} \leftarrow \pi_\theta^{\mathrm{syn}}(\mathcal{I}^{(\texttt{new})})$
        \If{$\mathrm{ValidateSchema}(\hat{\mathcal{S}}^{(\texttt{new})})$}
            \State $\mathcal{S} \leftarrow \mathcal{S} \cup \hat{\mathcal{S}}^{(\texttt{new})}$
            \State \textbf{break}
        \EndIf
        \State $\mathcal{I}^{(\texttt{new})} \leftarrow \mathrm{AppendSchemaFeedback}(\mathcal{I}^{(\texttt{new})})$
    \EndFor
\EndIf
\State \Return $\mathcal{S}$
\end{algorithmic}
\end{algorithm}

\begin{algorithm}[ht]
\caption{Version-upgrade protocol with Build+Fix regularization.}
\label{alg:protocol}
\begin{algorithmic}[1]
\Require Chain $v_0 \rightarrow v_1 \rightarrow \cdots \rightarrow v_N$; starting codebase $c_{v_0}$; specifications $\{\mathcal{S}_i\}_{i=1}^{N}$; agent $\pi_\theta$
\Ensure Agent-produced codebases $\{\hat{c}_{v_i}\}$ for all reached steps
\State $\hat{c}_{v_0} \leftarrow c_{v_0}$
\For{$i=1$ to $N$}
    \State $\hat{c}_{v_i} \leftarrow \pi_\theta(\mathcal{S}_i, \hat{c}_{v_{i-1}})$ \Comment{Build phase}
    \If{$\mathrm{BuildFailed}(\hat{c}_{v_i})$}
        \State \textbf{break} \Comment{terminate the entire chain}
    \EndIf
    \State $E_i \leftarrow \mathrm{RunHiddenTests}(\hat{c}_{v_i}, v_i)$
    \If{$\mathrm{HasExecutionErrors}(E_i)$} \Comment{Build+Fix: one controlled repair}
        \State $\rho_i \leftarrow \mathrm{WriteErrorReport}(E_i)$ \Comment{unique execution-error blocks}
        \State $\hat{c}_{v_i} \leftarrow \pi_\theta(\mathcal{S}_i, \rho_i, \hat{c}_{v_i})$ \Comment{Fix phase}
        \State $E_i \leftarrow \mathrm{RunHiddenTests}(\hat{c}_{v_i}, v_i)$
    \EndIf
\EndFor
\State \Return $\{\hat{c}_{v_i}\}$
\end{algorithmic}
\end{algorithm}

\clearpage
\section{Prompts}
\label{app:prompt}
\newcommand{\promptarg}[1]{\textcolor{blue}{\{#1\}}}
\newcommand{\omittedtag}[1]{\textcolor{gray}{#1}}

\subsection{Prompts for \synthesis}
\label{app:pro:synthesis}
This section shows the prompts used for \synthesis in \S\ref{sec:met:synthesis} where blue-highlighted text denotes the input parameters:
\begin{itemize}
    \item Prompt~\ref{prompt:matching} requests the agent to assign each diff hunk in a batch to one or more release-note tasks, \texttt{doc}, or \texttt{others}.
    \item Prompt~\ref{prompt:synthesis} requests the agent to synthesize an upgrade requirement for the target task using its matched code diffs, with the matched test diffs as supplementary reference only.
    Prompt~\ref{prompt:synthesis_cont} further requires the agent to output it in structured JSON format that includes four core parts: (1) requirements, (2) conceptual and grounded constraints and exceptions, (3) behaviors, and (4) acceptance criteria.
    \item Prompt~\ref{prompt:complete} requests the agent to review hunks assigned to \texttt{others} and synthesize additional tasks when they indicate meaningful behavioral changes not covered by the release note.
\end{itemize}

\begin{prompt}[ht]
\begin{AIBox}{Diff Hunk-to-Task Matching Prompt}
{\ttfamily\scriptsize
The current directory contains:\\
- \promptarg{release\_note\_file} (Release Note — each task has a task\_id and GitHub context)\\
- \promptarg{code\_diff\_file} (Code Diff)\\
- output.json (pre-filled with all task\_id keys as empty lists, plus "doc" and "others")\\
\\
Analyze the library update between two versions of a codebase, based on the Release Note and the CodeDiff. Map each diff hunk to the task\_id(s) it supports. Write the result to output.json.\\
\\
Rules:\\
- A hunk is identified by its @@ header up to and including the closing @@. Do NOT include trailing context (e.g., "@@ -185,6 +185,15 @@" not "@@ -185,6 +185,15 @@ class Foo:").\\
- A hunk can match multiple tasks (multi-label).\\
- "doc": purely documentation/comments/type-only changes with no runtime effect.\\
- "others": does not match any task.\\
- Every hunk must be assigned. Do NOT leave any unmatched.\\
\\
Output format — each task\_id maps to a list of:\\
\{"header": "diff --git a/... b/...", "hunks": ["@@ -a,b +c,d @@", ...]\}\\
}
\end{AIBox}
\caption{Prompt used to assign each code-diff hunk to one or more release-note tasks, \texttt{doc}, or \texttt{others}.}
\label{prompt:matching}
\end{prompt}

\begin{prompt}[ht]
\begin{AIBox}{Task Specification Synthesis Prompt}
{\ttfamily\scriptsize
The current directory contains:\\
- \promptarg{release\_note\_file} (Release Note content for task \promptarg{task\_id})\\
- \promptarg{code\_diff\_file} (Code Diff relevant to this task --- your primary focus)\\
- \promptarg{test\_diff\_file} (Test Diff for reference only)\\
- output.json (write your result here)\\
\\
Analyze the library update from \promptarg{old\_version} to \promptarg{new\_version} based on the Release Note and Code Diff. Synthesize a user-facing task requirement and write the result as JSON to output.json.\\
\\
\#\#\# Important guidelines\\
- The Test Diff is provided only as supplementary context to help you understand expected behavior. Do NOT derive requirements from test changes.\\
- The synthesized requirement MUST NOT include any test case modifications, test file changes, or testing instructions.\\
- The requirement must be feasible to understand and implement solely from the release note and code diff.\\
\\
\#\#\# Objectives\\
1. Understand what changed between versions from a user and product perspective\\
2. Produce a detailed breakdown of meaningful changes as a single upgrading task and requirement\\
3. Capture both the user requirement (from release note) and behavioral semantics (from code diff)\\
\\
\#\#\# Requirement fields (grounded vs conceptual)\\
1. \textbf{Grounded Specification}:\\
  - Grounded in release note and code diff evidence\\
  - May include precise public API names, modules, functions, classes, error types, and edge-case triggers\\
  - DO NOT include exact code snippets, messages, or implementation details\\
  - MUST NOT include patches, step-by-step implementation instructions, or exact code edits\\
\\
2. \textbf{Conceptual Specification}:\\
  - High-level user-facing, pure natural language specification\\
  - Focus on observable behavior contracts and acceptance criteria only\\
  - MUST NOT include localization cues (file paths, internal identifiers, diff-derived details)\\
  - DO NOT prescribe implementation strategies\\
\\
\#\#\# Field definitions\\
- \textasciigrave problem\_statement\textasciigrave: User-visible problem/feature. Checklist-style for fixes ("Ensure that ..."). No file paths or implementation steps.\\
- \textasciigrave expectation.grounded\textasciigrave: May mention public API names and observable signals (return/raise/warn/state). No code snippets, patches, file paths, exact messages.\\
- \textasciigrave expectation.conceptual\textasciigrave: High-level behavior description. May mention public API names, return/raise/warn types. No file paths, internal identifiers, patches.\\
- \textasciigrave constraints.grounded\textasciigrave: Non-regression / safety boundaries in precise terms. No "must use X / place at Y".\\
- \textasciigrave constraints.conceptual\textasciigrave: Same boundaries in high-level terms.\\
- \textasciigrave behavior\textasciigrave: List precise behavioral changes and observable signals. Derived from code diff, not test diff.\\
- \textasciigrave acceptance\_criteria\textasciigrave: Checklist items as observable checks (Given/When/Then or "When..., it should..."). Must describe production behavior, not test assertions.\\
}
\end{AIBox}
\caption{Prompt used to synthesize a user-facing upgrade task specification from the release note and task-relevant code diff.}
\label{prompt:synthesis}
\end{prompt}

\begin{prompt}[ht]
\begin{AIBox}{Task Specification Synthesis Prompt (cont.)}
{\ttfamily\scriptsize
\#\# Output\\
Return ONLY valid JSON wrapped in a markdown code block (\textasciigrave\textasciigrave\textasciigrave json ... \textasciigrave\textasciigrave\textasciigrave).\\
Output Format:\\
\textasciigrave\textasciigrave\textasciigrave json\\
\{\\
\hspace*{2em}"task\_id": "\promptarg{task\_id}",\\
\hspace*{2em}"title": "A short title of the task",\\
\hspace*{2em}"type": "select one from FEATURE | FIX | DEPENDENCY | BREAKING | TYPING | DOCUMENTATION | PERFORMANCE | REFACTOR | OTHER",\\
\hspace*{2em}"runtime\_impact": true | false,\\
\hspace*{2em}"description": "Description of what changed and why without implementation details.",\\
\hspace*{2em}"synthesized\_requirement": \{\\
\hspace*{4em}"problem\_statement": "...",\\
\hspace*{4em}"expectation": \{\\
\hspace*{6em}"grounded": "...",\\
\hspace*{6em}"conceptual": "..."\\
\hspace*{4em}\},\\
\hspace*{4em}"constraints": \{\\
\hspace*{6em}"grounded": "...",\\
\hspace*{6em}"conceptual": "..."\\
\hspace*{4em}\},\\
\hspace*{4em}"behavior": [\\
\hspace*{6em}"Precise behavioral changes: 'When X, function Y raises/returns/emits Z'",\\
\hspace*{6em}"Observable signals: 'Add a module-level variable that indicates X'",\\
\hspace*{6em}"..."\\
\hspace*{4em}],\\
\hspace*{4em}"acceptance\_criteria": ["..."]\\
\hspace*{2em}\},\\
\hspace*{2em}"significance": "high | medium | low",\\
\hspace*{2em}"confidence": "high | medium | low",\\
\hspace*{2em}"difficulty": "hard | medium | easy"\\
\}\\
}
\end{AIBox}
\caption{Continuation of the task specification synthesis prompt.}
\label{prompt:synthesis_cont}
\end{prompt}

\begin{prompt}[ht]
\begin{AIBox}{Task Completing Prompt}
{\ttfamily\scriptsize
The current directory contains:\\
- synthesized\_tasks.json (Already synthesized specifications from the release note)\\
- \promptarg{code\_diff\_file} (Unmatched code diff hunks not assigned to any task)\\
- \promptarg{test\_diff\_file} (Unmatched test diff hunks; supporting evidence only, may be empty)\\
- new\_tasks.json (Write your result here)\\
\\
Review the unmatched diff hunks from a library upgrade (\promptarg{old\_version} to \promptarg{new\_version}). Determine if any represent meaningful behavioral changes not already covered by synthesized\_tasks.json.\\
\\
Use \promptarg{code\_diff\_file} as the primary source of truth. Any new task you synthesize must be grounded in the code diff.\\
Use \promptarg{test\_diff\_file} only as supporting evidence to clarify or validate behavior suggested by the code diff.\\
Do not create a new task based solely on tests.\\
\\
If yes, write new task specifications to new\_tasks.json using the same schema as synthesized\_tasks.json, with IDs starting from "new\_task\_1".\\
If no, leave new\_tasks.json as `{}`.\\
Highlight the type of change for each new task from: "FEATURE | FIX | DEPENDENCY | BREAKING | TYPING | DOCUMENTATION | PERFORMANCE | REFACTOR | OTHER"\\

Ignore version bumps, formatting/linting, import reordering, and internal refactors with no observable behavior change.\\
}
\end{AIBox}
\caption{Prompt used for task-wise specification synthesis. Blue text indicates input arguments.}
\label{prompt:complete}
\end{prompt}

\clearpage
\subsection{Prompts for Evaluation}
\label{app:pro:evaluation}

This section shows the prompts used for the evaluation protocol in \S\ref{sec:exp:model}, where blue-highlighted text denotes the input parameters:
\begin{itemize}
    \item Prompt~\ref{prompt:build} is used in the Build stage, where the agent implements the target version upgrade from the synthesized specification.
    \item Prompt~\ref{prompt:fix} is used in the Fix stage, where the agent receives an execution error report and performs one controlled repair attempt.
\end{itemize}

\begin{prompt}[ht]
\begin{AIBox}{Upgrade Implementation (Build) Prompt}
{\ttfamily\scriptsize
Now implement the upgrade of \promptarg{package\_name} from version \promptarg{old\_version} to \promptarg{new\_version}.\\
The specification is at `./\promptarg{specs\_filename}`. The source code is under `./code/`. The original test suite has been explicitly removed.\\
Implement all required changes according to the specification. Ensure consistency across all files, especially when renaming or removing symbols and APIs.\\
You may also update ancillary version-bookkeeping files (e.g., `\_\_version\_\_` and CHANGELOG) when it matters for the upgrade.\\
Do NOT fetch, download, or request any external resource. Do NOT use the internet, WebSearch, WebFetch, remote APIs, package registries, or remote git hosts.
}
\end{AIBox}
\caption{Prompt used to instruct the agent to implement a package version upgrade from the old version to the target version.}
\label{prompt:build}
\end{prompt}

\begin{prompt}[ht]
\begin{AIBox}{Interface Repair (Fix) Prompt}
{\ttfamily\scriptsize
We ran the hidden test suite against your \promptarg{package\_name} upgrade (\promptarg{old\_version} to \promptarg{new\_version}) and found interface mismatches.\\
Read `./error\_report.md` for the list of issues and reconcile your public API names and signatures in `./code/` so that downstream callers resolve correctly.\\
The upgrade specification is still at `./\promptarg{specs\_filename}`.\\
Do NOT fetch, download, or request any external resource. Do NOT use the internet, WebSearch, WebFetch, remote APIs, package registries, or remote git hosts.
}
\end{AIBox}
\caption{Prompt used in the Build+Fix regularization stage, where the agent receives an error report after hidden-test execution.}
\label{prompt:fix}
\end{prompt}

\clearpage
\section{Build+Fix Trajectories}
\label{app:trajectories}
\begin{figure}[ht]
    \centering
    \includegraphics[width=0.7\linewidth]{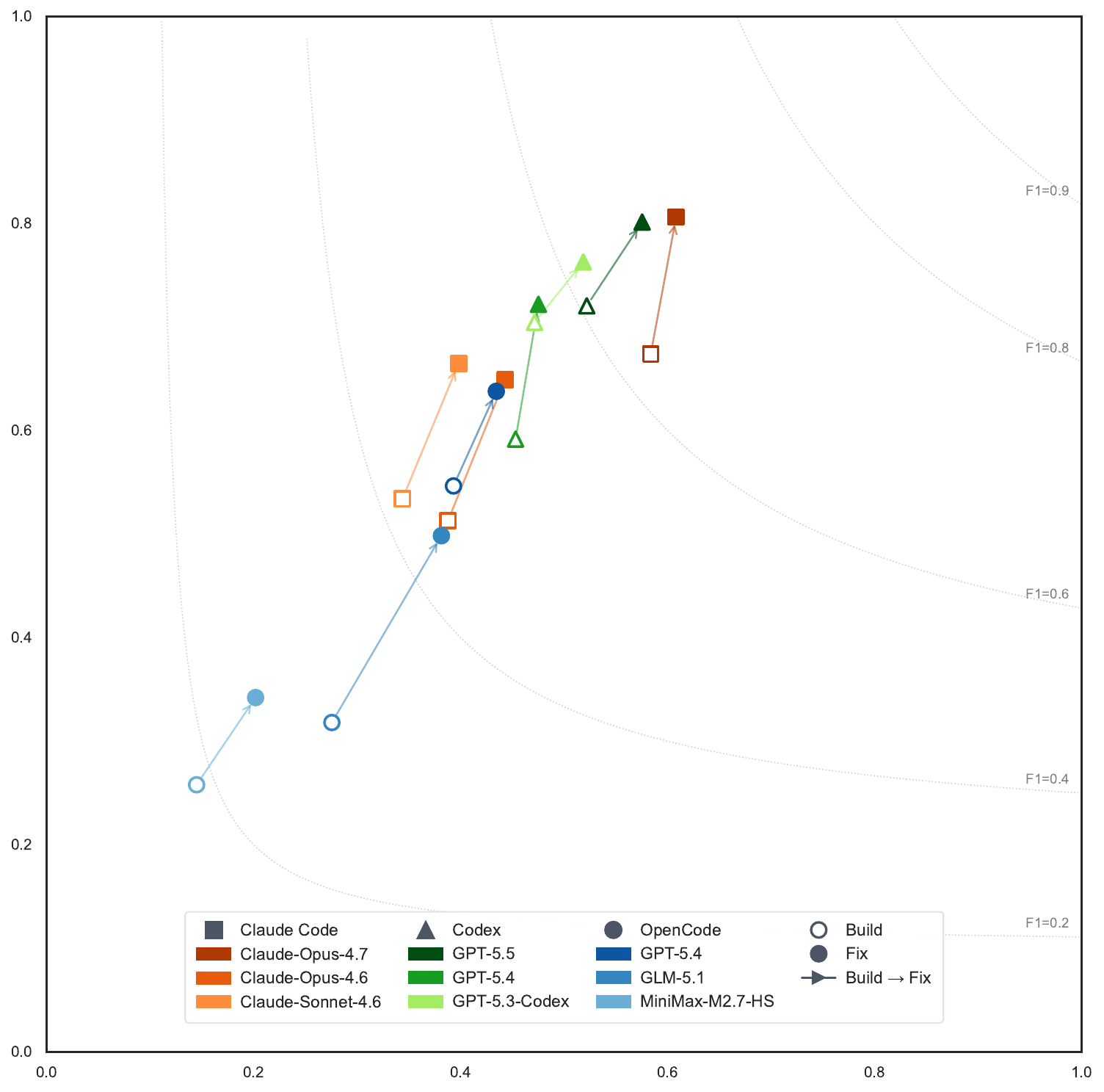}
    \caption{Effect of Build+Fix on resolving rate and precision. Hollow and filled markers denote Build and Build+Fix, respectively. Gray dashed curves indicate F1-score iso-curves.}
    \label{fig:build_to_fix}
\end{figure}

To complement the effects of the Build+Fix regime in  \S\ref{sec:res:overall} and Table~\ref{tab:avg_results}, we visualize the trajectory from Build to Build+Fix in the resolving-precision plane in Figure~\ref{fig:build_to_fix}.
Most agents show a larger upward than rightward shift, indicating that Build+Fix improves precision more than resolving rate.
The figure also shows that open-weight models remain below closed-source models even after Build+Fix, suggesting that the repair step does not erase the underlying capability gap.

\clearpage
\section{Cross-Chain Precision and F1}
\label{app:heatmap}
\begin{figure}[ht]
    \centering
    \includegraphics[width=1.0\linewidth]{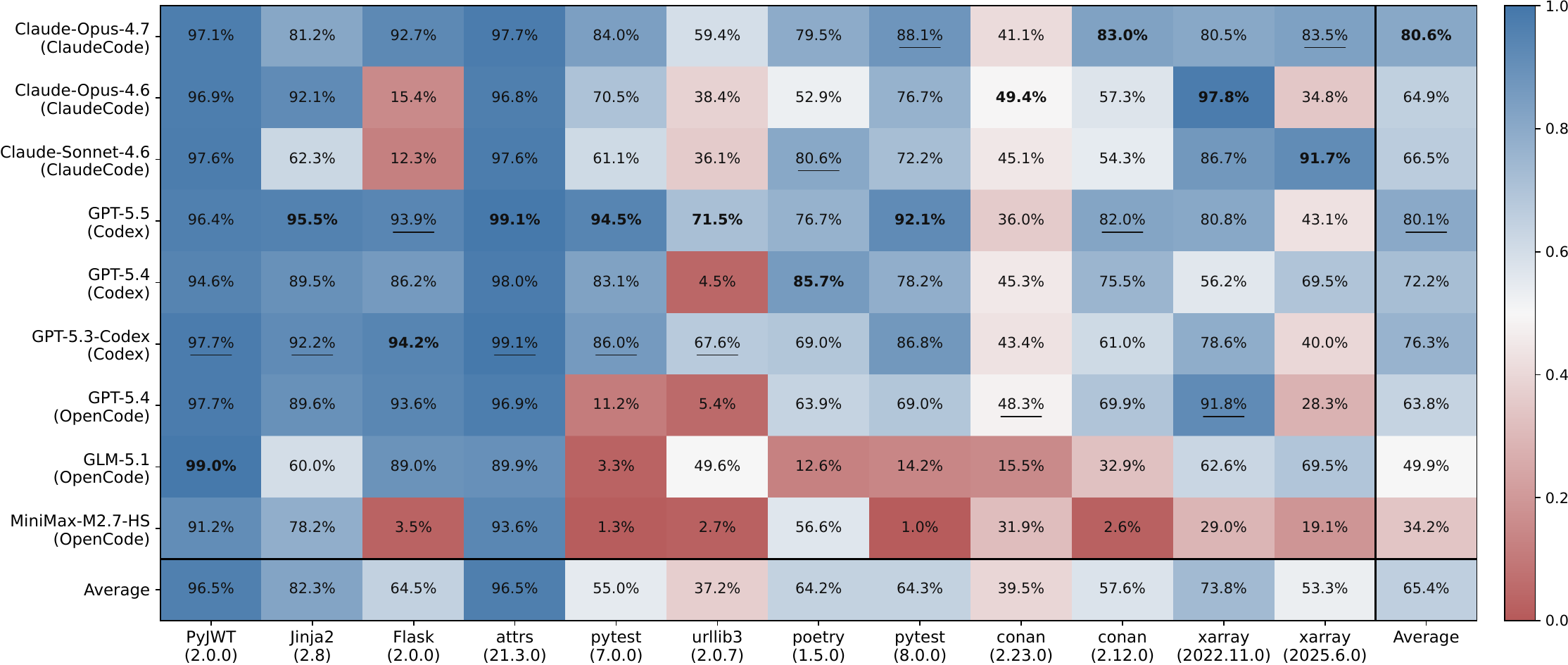}
    \caption{Build+Fix precision by agent and chain, order left-to-right by increasing difficulty.}
    \label{fig:heatmap_precision}
\end{figure}

\begin{figure}[ht]
    \centering
    \includegraphics[width=1.0\linewidth]{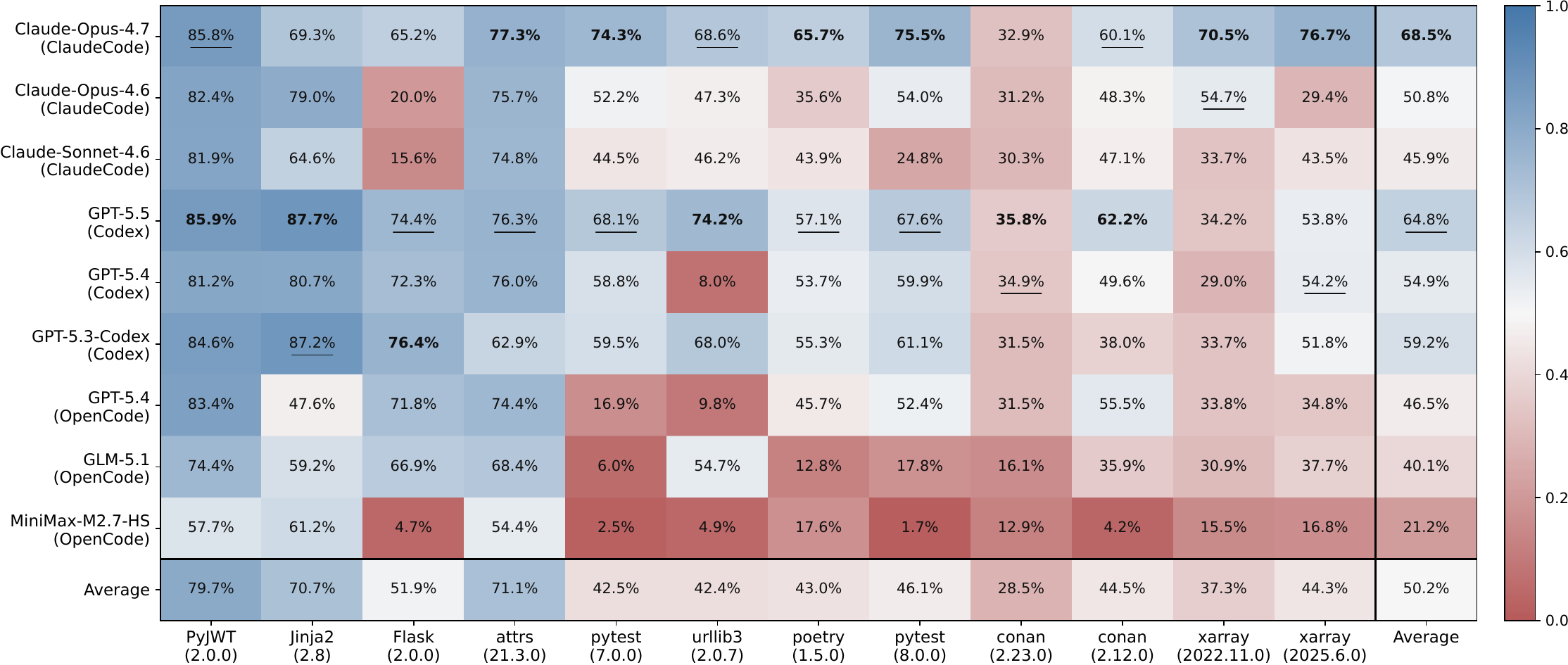}
    \caption{Build+Fix F1 by agent and chain, order left-to-right by increasing difficulty.}
    \label{fig:heatmap_f1}
\end{figure}

To complement the heatmap Figure~\ref{fig:heatmap_resolving} in \S\ref{sec:res:cross}, we report the corresponding Build+Fix precision and F1 heatmaps in Figures~\ref{fig:heatmap_precision} and~\ref{fig:heatmap_f1}, respectively.
Overall, precision is substantially higher than resolving rate across most chains, which raises the resulting F1 scores.
This suggests that agents are often relatively targeted when they implement upgrade-specific behaviors.
However, chains such as \texttt{urllib3} and \texttt{conan} 2.23.0 also show lower precision, which further lowers their F1 scores.
Hence, the F1 heatmap provides a balanced view of both chain difficulty and maintenance capability, while the resolving-rate heatmap more directly reflects the requirement implementation ability.


\section{Per-Step Resolving Trajectories}
\label{app:perstep}

\begin{figure}[ht]
    \centering
    \includegraphics[width=\linewidth]{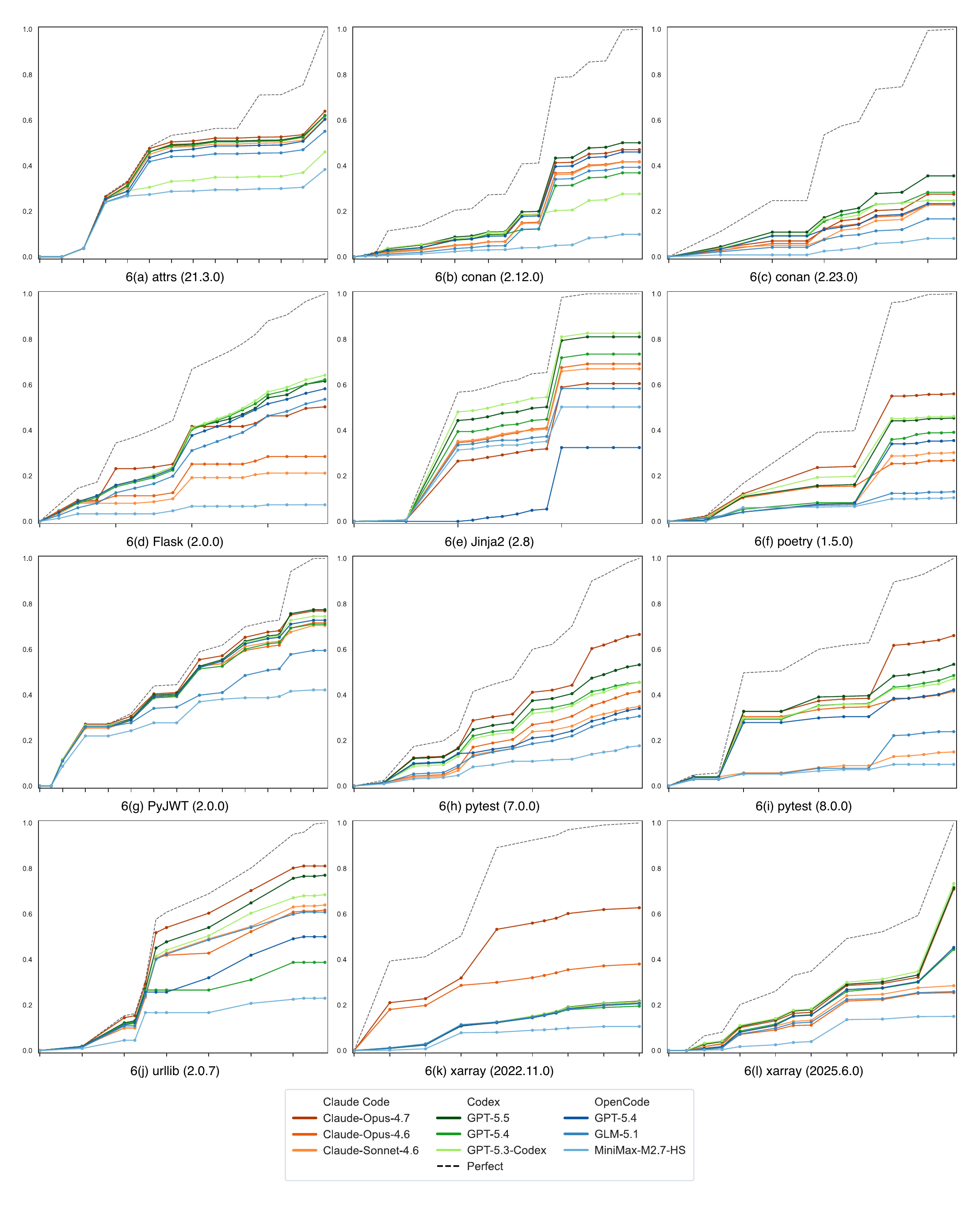}
    \caption{Per-step Build+Fix resolving-progress trajectories across all upgrade chains.
    Each curve point corresponds to one version-upgrade step.
    Note that the x-axis ticks mark only minor or major version milestones and are spaced proportionally by semantic-version distance.}
    \label{fig:recall_curve}
\end{figure}

To visualize how agents progress along each upgrade chain, we plot the cumulative resolving progress at each upgrade for all chains in Figure~\ref{fig:recall_curve}.
For a chain of length $n$, the value at step $x$ is defined as:
\begin{equation*}
\mathrm{Resolving}_{\le x} = \frac{\sum_{i=1}^{x} \mathrm{TP}_i}{\sum_{i=1}^{n}(\mathrm{TP}_i+\mathrm{FN}_i)}.
\end{equation*}
The figure shows that agents usually progress smoothly across patch-level upgrades, where the number of upgrade-specific tests and the scope of changes is smaller.
In contrast, larger separations between agents often emerge around minor or major upgrades, where the upgrade scope becomes broader and missed requirements accumulate more visibly.
% This supports that the difficulty of \methodname comes not only from individual upgrade steps, but also from sustaining progress across increasingly demanding version transitions.

\clearpage
\section{Examples}
\label{app:examples}

This appendix uses the \texttt{pytest} 8.2.0 $\rightarrow$ 8.2.1 upgrade as an example to illustrate the \methodname pipeline in \S\ref{sec:methodology} and the resulting upgrade specification used in \S\ref{sec:results}.

\subsection{\synthesis Examples}
\label{app:exa:synthesis}
As described in \S\ref{sec:met:benchmark}, we first collect the raw release note, shown in Example~\ref{example:raw-release-note}, and decompose it into task-structured entries with the corresponding GitHub issues or pull requests, shown in Example~\ref{example:task-structured-release-note}.
Then, as described in \S\ref{sec:met:synthesis}, the agent takes the task-structured release note and code diffs, performs hunk-to-task matching, and produces the intermediate mapping document in Example~\ref{example:hunk-mapping-result}.
Next, the agent synthesizes a structured task specification for each matched release-note task, as shown in Example~\ref{example:synthesized-task-spec}.
Finally, we select the required fields and integrate all task-level specifications into the final upgrade specification for \texttt{pytest} ver. 8.2.0 $\rightarrow$ 8.2.1, shown in Example~\ref{example:granularity-l3}.

\begin{example}[ht]
\begin{AIBox}{Raw Release Note (\texttt{.md})}
{\ttfamily\scriptsize
\#\# pytest 8.2.1 (2024-05-19)[¶](\#pytest-8-2-1-2024-05-19)\\
\#\#\# Improvements[¶](\#improvements)\\
- [\#12334](https://github.com/pytest-dev/pytest/issues/12334): Support for Python 3.13 (beta1 at the time of writing).\\
\\
\#\#\# Bug Fixes[¶](\#id254)\\
- [\#12120](https://github.com/pytest-dev/pytest/issues/12120): Fix `PermissionError` crashes arising from directories which are not selected on the command-line.\\
- [\#12191](https://github.com/pytest-dev/pytest/issues/12191): Keyboard interrupts and system exits are now properly handled during the test collection.\\
- [\#12300](https://github.com/pytest-dev/pytest/issues/12300): Fixed handling of ‘Function not implemented’ error under squashfuse\_ll, which is a different way to say that the mountpoint is read-only.\\
- [\#12308](https://github.com/pytest-dev/pytest/issues/12308): Fix a regression in pytest 8.2.0 where the permissions of automatically-created `.pytest\_cache` directories became `rwx------` instead of the expected `rwxr-xr-x`.\\
\\
\#\#\# Trivial/Internal Changes[¶](\#trivial-internal-changes)\\
- [\#12333](https://github.com/pytest-dev/pytest/issues/12333): pytest releases are now attested using the [Artifact Attestation](https://github.blog/2024-05-02-introducing-artifact-attestations-now-in-public-beta/) support from GitHub, allowing users to verify the provenance of pytest’s sdist and wheel artifacts.
}
\end{AIBox}
\caption{Maintainer release note for the pytest 8.2.1 upgrade.}
\label{example:raw-release-note}
\end{example}
\clearpage

\begin{example}[ht]
\begin{AIBox}{Task-Structured Release Note (\texttt{.json})}
{\ttfamily\scriptsize
[\{\\
\hspace*{2em}"task\_id": "task\_1",\\
\hspace*{2em}"content": "[\#12334] Support for Python 3.13 (beta1).",\\
\hspace*{2em}"github": [\{\\
\hspace*{3em}"link": "https://github.com/pytest-dev/pytest/pull/12334",\\
\hspace*{3em}"type": "pr",\\
\hspace*{3em}"content": "\#\#\# GitHub Pull Request \#12334 Add Python 3.13 support\textbackslash{}nFix \#12323"\\
\hspace*{2em}\}]\},\{\\
\hspace*{2em}"task\_id": "task\_2",\\
\hspace*{2em}"content": "[\#12120] Fix PermissionError crashes from unselected directories.",\\
\hspace*{2em}"github": [\{\\
\hspace*{3em}"link": "https://github.com/pytest-dev/pytest/issues/12120",\\
\hspace*{3em}"type": "issue",\\
\hspace*{3em}"content": "\#\#\# GitHub Issue \#12120 Pytest crashes if a subdirectory has no read access\textbackslash{}nPytest crashes if there is a subdirectory it does not \omittedtag{[omitted...]}"\\
\hspace*{2em}\}]\},\{\\
\hspace*{2em}"task\_id": "task\_3",\\
\hspace*{2em}"content": "[\#12191] KeyboardInterrupt and SystemExit are handled during collection.",\\
\hspace*{2em}"github": [\{\\
\hspace*{3em}"link": "https://github.com/pytest-dev/pytest/pull/12191",\\
\hspace*{3em}"type": "pr",\\
\hspace*{3em}"content": "\#\#\# GitHub Pull Request \#12191 Consider KeyboardInterrupt/SystemExit at collection time\textbackslash{}n![Screenshot from 2024-04-06 21-48-46] \omittedtag{[omitted...]}"\\
\hspace*{2em}\}]\},\{\\
\hspace*{2em}"task\_id": "task\_4",\\
\hspace*{2em}"content": "[\#12300] Treat squashfuse\_ll ENOSYS as a read-only mountpoint case.",\\
\hspace*{2em}"github": [\{\\
\hspace*{3em}"link": "https://github.com/pytest-dev/pytest/issues/12300",\\
\hspace*{3em}"type": "issue",\\
\hspace*{3em}"content": "\#\#\# GitHub Issue \#12300 Crashing under a squashfuse\_ll read-only mount\textbackslash{}npytest is crashing with OSError: [Errno 38] Function not implemented \omittedtag{[omitted...]}"\\
\hspace*{2em}\}]\},\{\\
\hspace*{2em}"task\_id": "task\_5",\\
\hspace*{2em}"content": "[\#12308] Restore expected permissions for automatically-created .pytest\_cache directories.",\\
\hspace*{2em}"github": [\{\\
\hspace*{3em}"link": "https://github.com/pytest-dev/pytest/issues/12308",\\
\hspace*{3em}"type": "issue",\\
\hspace*{3em}"content": "\#\#\# GitHub Issue \#12308 EACCES: permission denied, scandir .pytest\_cache\textbackslash{}nWhen upgrading to Pytest 8.2.0, I was getting an error \omittedtag{[omitted...]}"\\
\hspace*{2em}\}]\},\{\\
\hspace*{2em}"task\_id": "task\_6",\\
\hspace*{2em}"content": "[\#12333] Attest pytest release artifacts with GitHub Artifact Attestations.",\\
\hspace*{2em}"github": [\{\\
\hspace*{3em}"link": "https://github.com/pytest-dev/pytest/pull/12333",\\
\hspace*{3em}"type": "pr",\\
\hspace*{3em}"content": "\#\#\# GitHub Pull Request \#12333 Attest package provenance\textbackslash{}nThis uses the new build provenance support added in build-and-inspect-python-package \omittedtag{[omitted...]}"\\
\hspace*{2em}\}]\}]\\
}
\end{AIBox}
\caption{Task-structured release-note entries with linked GitHub issue and pull-request context.}
\label{example:task-structured-release-note}
\end{example}
\clearpage

\begin{example}[ht]
\begin{AIBox}{Intermediate Hunk Mapping Result (\texttt{.json})}
{\ttfamily\scriptsize
\{\\
\hspace*{1em}"task\_1": [\{\\
\hspace*{2em}"diff\_header": "diff --git a/src/\_pytest/pytester.py b/src/\_pytest/pytester.py",\\
\hspace*{2em}"old\_path": "a/src/\_pytest/pytester.py",\\
\hspace*{2em}"new\_path": "b/src/\_pytest/pytester.py",\\
\hspace*{2em}"hunks": ["@@ -289,7 +289,8 @@", "@@ -760,6 +761,9 @@"]\\
\hspace*{1em}\}, \omittedtag{[omitted...]}],\\
\hspace*{1em}"task\_2": [\{\\
\hspace*{2em}"diff\_header": "diff --git a/src/\_pytest/python.py b/src/\_pytest/python.py",\\
\hspace*{2em}"old\_path": "a/src/\_pytest/python.py",\\
\hspace*{2em}"new\_path": "b/src/\_pytest/python.py",\\
\hspace*{2em}"hunks": ["@@ -176,7 +176,12 @@"]\\
\hspace*{1em}\}],\\
\hspace*{1em}"task\_3": [\{\\
\hspace*{2em}"diff\_header": "diff --git a/src/\_pytest/runner.py b/src/\_pytest/runner.py",\\
\hspace*{2em}"old\_path": "a/src/\_pytest/runner.py",\\
\hspace*{2em}"new\_path": "b/src/\_pytest/runner.py",\\
\hspace*{2em}"hunks": ["@@ -388,7 +388,9 @@"]\\
\hspace*{1em}\}],\\
\hspace*{1em}"task\_4": [\{\\
\hspace*{2em}"diff\_header": "diff --git a/src/\_pytest/assertion/rewrite.py b/src/\_pytest/assertion/rewrite.py",\\
\hspace*{2em}"old\_path": "a/src/\_pytest/assertion/rewrite.py",\\
\hspace*{2em}"new\_path": "b/src/\_pytest/assertion/rewrite.py",\\
\hspace*{2em}"hunks": ["@@ -1171,7 +1171,10 @@"]\\
\hspace*{1em}\}],\\
\hspace*{1em}"task\_5": [\{\\
\hspace*{2em}"diff\_header": "diff --git a/src/\_pytest/cacheprovider.py b/src/\_pytest/cacheprovider.py",\\
\hspace*{2em}"old\_path": "a/src/\_pytest/cacheprovider.py",\\
\hspace*{2em}"new\_path": "b/src/\_pytest/cacheprovider.py",\\
\hspace*{2em}"hunks": ["@@ -213,6 +213,13 @@"]\\
\hspace*{1em}\}],\\
\hspace*{1em}"task\_6": [],\\
\hspace*{1em}"doc": [\{\\
\hspace*{2em}"diff\_header": "diff --git a/src/\_pytest/outcomes.py b/src/\_pytest/outcomes.py",\\
\hspace*{2em}"old\_path": "a/src/\_pytest/outcomes.py",\\
\hspace*{2em}"new\_path": "b/src/\_pytest/outcomes.py",\\
\hspace*{2em}"hunks": ["@@ -114,6 +114,9 @@", "@@ -142,6 +145,9 @@", "@@ -163,6 +169,9 @@", "@@ -188,6 +197,9 @@", "@@ -227,6 +239,9 @@"]\\
\hspace*{1em}\}, \omittedtag{[omitted...]}],\\
\hspace*{1em}"others": [\{\\
\hspace*{2em}"diff\_header": "diff --git a/src/\_pytest/config/\_\_init\_\_.py b/src/\_pytest/config/\_\_init\_\_.py",\\
\hspace*{2em}"old\_path": "a/src/\_pytest/config/\_\_init\_\_.py",\\
\hspace*{2em}"new\_path": "b/src/\_pytest/config/\_\_init\_\_.py",\\
\hspace*{2em}"hunks": ["@@ -574,8 +574,8 @@"]\\
\hspace*{1em}\}, \omittedtag{[omitted...]}]\\
\}\\
}
\end{AIBox}
\caption{Intermediate mapping from release-note tasks to matched code-diff hunks.}
\label{example:hunk-mapping-result}
\end{example}
\clearpage

\begin{example}[ht]
\begin{AIBox}{Synthesized Task Specification Output (\texttt{.json})}
{\ttfamily\scriptsize
\{\\
\hspace*{1em}"task\_2": \{\\
\hspace*{2em}"task\_id": "task\_2",\\
\hspace*{2em}"title": "Avoid collection crashes from unreadable unselected directories",\\
\hspace*{2em}"type": "FIX",\\
\hspace*{2em}"runtime\_impact": true,\\
\hspace*{2em}"description": "The update fixes a runtime collection failure in pytest when an unreadable directory exists in the project tree but is not part of the command-line selection. Instead of aborting collection with a PermissionError while checking whether that directory is a Python package, pytest should ignore that directory and continue collecting the explicitly requested tests.",\\
\hspace*{2em}"synthesized\_requirement": \{\\
\hspace*{3em}"problem\_statement": "Ensure that running pytest against selected test paths does not crash when unrelated directories in the project tree are unreadable.",\\
\hspace*{3em}"expectation": \{\\
\hspace*{4em}"grounded": "During directory collection, pytest should treat a PermissionError raised while checking whether a directory contains a package initializer as a non-collectable directory and continue processing the command-line-selected test targets. If the unreadable directory is not selected for collection, pytest should not abort the session because of that directory.",\\
\hspace*{4em}"conceptual": "Pytest should remain resilient to unreadable directories that are outside the user's requested test scope. Running selected tests should continue normally instead of failing because pytest inspected an unrelated directory it cannot read."\\
\hspace*{3em}\},\\
\hspace*{3em}"constraints": \{\\
\hspace*{4em}"grounded": "The fix is limited to PermissionError encountered while determining whether a directory should be collected as a Python package. Directories that can be inspected and contain an initializer should still be collected as packages, and directories without that condition should remain uncollected as before.",\\
\hspace*{4em}"conceptual": "Only the unreadable-directory crash should be removed. Existing package discovery behavior for accessible directories should remain unchanged."\\
\hspace*{3em}\},\\
\hspace*{3em}"behavior": [\\
\hspace*{4em}"When pytest inspects a directory during collection and reading its package initializer status raises PermissionError, pytest skips collecting that directory instead of propagating the exception.",\\
\hspace*{4em}"An unreadable directory that was not selected on the command line no longer prevents pytest from collecting and running the explicitly requested tests.",\\
\hspace*{4em}"Accessible directories that qualify as Python packages continue to be collected as package nodes."\\
\hspace*{3em}],\\
\hspace*{3em}"acceptance\_criteria": [\\
\hspace*{4em}"When a project contains an unreadable sibling directory outside the selected test path, invoking pytest on the selected test path should not fail with PermissionError caused by that sibling directory.",\\
\hspace*{4em}"When pytest encounters PermissionError while determining whether a directory is a package, it should treat that directory as not collectable and continue collection.",\\
\hspace*{4em}"When a directory is accessible and contains a valid package initializer, pytest should still collect it as a package."\\
\hspace*{3em}]\\
\hspace*{2em}\},\\
\hspace*{2em}"significance": "medium", "confidence": "high", "difficulty": "easy"\\
\hspace*{1em}\}\\
\}\\
}
\end{AIBox}
\caption{Structured synthesized requirement for one matched upgrade task.}
\label{example:synthesized-task-spec}
\end{example}
\clearpage

\begin{example}[ht]
\begin{AIBox}{Upgrade Specification (\texttt{.md})}
{\ttfamily\scriptsize
\# Upgrade Specification from 8.2.0 to 8.2.1\\
\#\# New Features\\
\#\#\# Task 1 Add Python 3.13 compatibility and stricter pytester.makefile validation\\
Ensure that the library works correctly on Python 3.13 for traceback and pytester workflows that depend on frame-local state, and ensure that invalid `Pytester.makefile()` calls with `ext=None` are rejected explicitly.\\
- Expectation: The library should support Python 3.13 without breaking traceback analysis or `pytester` helper workflows because of changes in how Python exposes local variables. Invalid file-creation requests that omit an extension value entirely should fail clearly\\ instead of being processed.\\
- Constraints: The compatibility fix should preserve prior observable behavior outside the Python 3.13-specific locals change. Normal file creation through the helper should continue to work, and only clearly invalid missing-extension usage should be blocked.\\
\\
\#\# Bug Fixes\\
\#\#\# Task 2 Avoid collection crashes from unreadable unselected directories\\
Ensure that running pytest against selected test paths does not crash when unrelated directories in the project tree are unreadable.\\
- Expectation: Pytest should remain resilient to unreadable directories that are outside the user's requested test scope. Running selected tests should continue normally instead of failing because pytest inspected an unrelated directory it cannot read.\\
- Constraints: Only the unreadable-directory crash should be removed. Existing package discovery behavior for accessible directories should remain unchanged.\\
\#\#\# Task 3 Preserve system-exit behavior during pytest collection\\
Ensure that pytest handles `KeyboardInterrupt` and `SystemExit` correctly when they occur during test collection, so the run stops immediately instead of treating them like normal collection errors.\\
- Expectation: If test discovery is interrupted by a user interrupt or an explicit process exit, pytest should stop discovery immediately and surface that interrupt or exit as such, rather than reporting it as a standard collection failure.\\
- Constraints: Only true system-level interruptions during discovery should get special handling. Normal collection results and ordinary collection errors should keep their existing behavior.\\
\#\#\# Task 4 Handle read-only squashfuse\_ll mounts without crashing\\
Ensure that running the library on a read-only squashfuse\_ll mount does not crash when cache directories cannot be created, and that this condition is treated the same way as other read-only filesystem failures.\\
- Expectation: When the runtime is on a read-only filesystem that reports the failure using 'Function not implemented' instead of a standard read-only error, the library should skip cache directory creation and continue operating instead of crashing.\\
- Constraints: Treat only recognized read-only mount failures as recoverable, and do not hide unrelated filesystem errors.\\
\#\#\# Task 5 Restore default .pytest\_cache directory permissions\\
Ensure that automatically created .pytest\_cache directories no longer use owner-only permissions after upgrading to pytest 8.2.1, and instead remain accessible according to normal system defaults so repository-scanning tools do not fail with permission errors.\\
- Expectation: Pytest-created cache directories should behave like ordinary project directories: accessible to other tools according to the environment's normal permission policy, instead of becoming unexpectedly private to the current user.\\
- Constraints: The change should remove the regression without making cache directory permissions broader than the system would normally allow, and without changing how users interact with pytest's cache feature.\\
\\
\#\# Additional Changes\\
- Ensure that users upgrading to pytest 8.2.1 can verify the provenance of official pytest distribution artifacts, including both source distributions and wheels, without changing pytest’s runtime behavior.\\
}
\end{AIBox}
\caption{Core specification used in the main experiments in \S\ref{sec:res:overall}.}
\label{example:granularity-l3}
\end{example}
\clearpage

\subsection{Specification Granularity Examples}
\label{app:exa:granularity}

In \S\ref{sec:res:granularity}, we study how specification granularity affects agent performance.
We present the five specification variants used in the granularity study:
\begin{itemize}[leftmargin=*]
    \item \textbf{L1}: Raw release notes with GitHub issues and pull requests, shown in Example~\ref{example:granularity-l1}.
    \item \textbf{L2}: Problem statements only, shown in Example~\ref{example:granularity-l2}.
    \item \textbf{L3}: Problem statements with conceptual expectations and constraints, shown in Example~\ref{example:granularity-l3}.
    \item \textbf{L4}: Problem statements with grounded expectations and constraints, shown in Example~\ref{example:granularity-l4}.
    \item \textbf{L5}: Problem statements, grounded expectations and constraints, behaviors, and acceptance criteria, shown in Example~\ref{example:granularity-l5}.
\end{itemize}

\begin{example}[ht]
\begin{AIBox}{L2: Problem-Only Specification (\texttt{.md})}
{\ttfamily\scriptsize
\# Upgrade Specification from 8.2.0 to 8.2.1\\
\#\# New Features\\
\#\#\# Task 1 Add Python 3.13 compatibility and stricter pytester.makefile validation\\
Ensure that the library works correctly on Python 3.13 for traceback and pytester workflows that depend on frame-local state, and ensure that invalid `Pytester.makefile()` calls with `ext=None` are rejected explicitly.\\
\\
\#\# Bug Fixes\\
\#\#\# Task 2 Avoid collection crashes from unreadable unselected directories\\
Ensure that running pytest against selected test paths does not crash when unrelated directories in the project tree are unreadable.\\
\\
\#\#\# Task 3 Preserve system-exit behavior during pytest collection\\
Ensure that pytest handles `KeyboardInterrupt` and `SystemExit` correctly when they occur during test collection, so the run stops immediately instead of treating them like normal collection errors.\\
\\
\#\#\# Task 4 Handle read-only squashfuse\_ll mounts without crashing\\
Ensure that running the library on a read-only squashfuse\_ll mount does not crash when cache directories cannot be created, and that this condition is treated the same way as other read-only filesystem failures.\\
\\
\#\#\# Task 5 Restore default .pytest\_cache directory permissions\\
Ensure that automatically created .pytest\_cache directories no longer use owner-only permissions after upgrading to pytest 8.2.1, and instead remain accessible according to normal system defaults so repository-scanning tools do not fail with permission errors.\\
\\
\#\# Additional Changes\\
- Ensure that users upgrading to pytest 8.2.1 can verify the provenance of official pytest distribution artifacts, including both source distributions and wheels, without changing pytest’s runtime behavior.
}
\end{AIBox}
\caption{L2 specification: problem statements only.}
\label{example:granularity-l2}
\end{example}
\clearpage

\begin{example}[ht]
\begin{AIBox}{Specification L1: Specification with Raw Release Notes and GitHub Context (\texttt{.md})}
{\ttfamily\scriptsize
1. [\#12334](https://github.com/pytest-dev/pytest/issues/12334): Support for Python 3.13 (beta1 at the time of writing).\\
2. [\#12120](https://github.com/pytest-dev/pytest/issues/12120): Fix `PermissionError` crashes arising from directories which are not selected on the command-line.\\
3. [\#12191](https://github.com/pytest-dev/pytest/issues/12191): Keyboard interrupts and system exits are now properly handled during the test collection.\\
4. [\#12300](https://github.com/pytest-dev/pytest/issues/12300): Fixed handling of `Function not implemented' error under squashfuse\_ll, which is a different way to say that the mountpoint is read-only.\\
5. [\#12308](https://github.com/pytest-dev/pytest/issues/12308): Fix a regression in pytest 8.2.0 where automatically-created `.pytest\_cache` directories became `rwx------` instead of `rwxr-xr-x`.\\
6. [\#12333](https://github.com/pytest-dev/pytest/issues/12333): pytest releases are now attested using GitHub Artifact Attestations, allowing users to verify sdist and wheel provenance.\\
\\
\#\#\# GitHub Pull Request \#12334 Add Python 3.13 support\\
Fix \#12323\\
\#\#\# GitHub Issue \#12120 Pytest crashes if a subdirectory has no read access\\
Pytest crashes if there is a subdirectory it does not \omittedtag{[omitted...]}\\
\#\#\# GitHub Pull Request \#12191 Consider KeyboardInterrupt/SystemExit at collection time\\
![Screenshot from 2024-04-06 21-48-46](https://github.com/pytest-dev/pytest/assets/166057949/8c8ae3b7-22bb-\\4297-a1e0-6d03c474bab4)\\
\textless{}!--\\
Thanks for submitting a PR, your contribution is really appreciated!\\
Here is a quick checklist that should be present in PRs.\\
- [ ] Include documentation when adding new features.\\
- [ ] Include new tests or update existing tests when applicable.\\
- [X] Allow maintainers to push and squash when merging my commits. Please uncheck this if you prefer to squash the commits yourself.\\
If this change fixes an issue, please:\\
- [ ] Add text like ``closes \#XYZW`` to the PR description and/or commits (where ``XYZW`` is the issue number). See the [github docs](https://help.github.com/en/github/managing-your-work-on-github/linking-a-pull-request-to-an-issue\#\\linking-a-pull-request-to-an-issue-using-a-keyword) for more information.\\
Unless your change is trivial or a small documentation fix (e.g., a typo or reword of a small section) please:\\
- [ ] Create a new changelog file in the `changelog` folder, with a name like `\textless{}ISSUE NUMBER\textgreater{}.\textless{}TYPE\textgreater{}.rst`. See [changelog/README.rst](https://github.com/pytest-dev/pytest/blob/main/changelog/README.rst) for details.\\
\hspace*{1em}Write sentences in the **past or present tense**, examples:\\
\hspace*{1em}* *Improved verbose diff output with sequences.*\\
\hspace*{1em}* *Terminal summary statistics now use multiple colors.*\\
\hspace*{1em}Also make sure to end the sentence with a `.'.\\
- [ ] Add yourself to `AUTHORS` in alphabetical order.\\
--\textgreater{}\\
\#\#\# GitHub Issue \#12300 Crashing under a squashfuse\_ll read-only mount\\
pytest is crashing with OSError: [Errno 38] Function not implemented \omittedtag{[omitted...]}\\
\#\#\# GitHub Issue \#12308 EACCES: permission denied, scandir .pytest\_cache\\
When upgrading to Pytest 8.2.0, I was getting an error \omittedtag{[omitted...]}\\
\#\#\# GitHub Pull Request \#12333 Attest package provenance\\
This uses the new build provenance support added in build-and-inspect-python-package \omittedtag{[omitted...]}\\
}
\end{AIBox}
\caption{L1 specification: release notes with linked GitHub issues and pull requests.}
\label{example:granularity-l1}
\end{example}
\clearpage

\begin{example}[ht]
\begin{AIBox}{L4: Grounded Expectation Specification (\texttt{.md})}
{\ttfamily\scriptsize
\# Upgrade Specification from 8.2.0 to 8.2.1\\
\#\# New Features\\
\#\#\# Task 1 Add Python 3.13 compatibility and stricter pytester.makefile validation\\
Ensure that the library works correctly on Python 3.13 for traceback and pytester workflows that depend on frame-local state, and ensure that invalid `Pytester.makefile()` calls with `ext=None` are rejected explicitly.\\
- Expectation: On Python 3.13, behavior that reads frame locals in traceback processing and in `pytester` hook-recording checks must continue to work even when frame locals are exposed through a proxy-style mapping rather than a plain dict. `Pytester.makefile()` must raise\\ `TypeError` when `ext` is `None` instead of accepting that input.\\
- Constraints: Existing behavior for supported Python versions must remain intact. Recursion detection in tracebacks must still treat repeated frames with equivalent local-variable state as recursive. Valid `Pytester.makefile()` calls with a real extension string must continue\\ to follow the existing extension rules, with only `None` being newly rejected.\\
\\
\#\# Bug Fixes\\
\#\#\# Task 2 Avoid collection crashes from unreadable unselected directories\\
Ensure that running pytest against selected test paths does not crash when unrelated directories in the project tree are unreadable.\\
- Expectation: During collection, pytest should treat a `PermissionError` while checking package-initializer status as a non-collectable directory and continue selected test collection. \omittedtag{[omitted...]}\\
- Constraints: The fix is limited to `PermissionError` during package-discovery checks; accessible package directories should keep existing collection behavior. \omittedtag{[omitted...]}\\
\\
\#\#\# Task 3 Preserve system-exit behavior during pytest collection\\
Ensure that pytest handles `KeyboardInterrupt` and `SystemExit` correctly during test collection.\\
- Expectation: Collection-time `KeyboardInterrupt` and `SystemExit` should be re-raised as system exceptions rather than converted into ordinary collection failures. \omittedtag{[omitted...]}\\
- Constraints: Other collection outcomes and non-system exceptions should continue to follow existing collection reporting semantics.\\
\\
\#\#\# Task 4 Handle read-only squashfuse\_ll mounts without crashing\\
Ensure that running on a read-only squashfuse\_ll mount does not crash when cache directories cannot be created.\\
- Expectation: Assertion rewrite cache setup should treat `errno.ENOSYS` from squashfuse\_ll as a recoverable read-only filesystem condition, like `errno.EROFS`. \omittedtag{[omitted...]}\\
- Constraints: Only recognized read-only-equivalent cache-directory failures should be suppressed; unrelated `OSError` values must still propagate.\\
\\
\#\#\# Task 5 Restore default .pytest\_cache directory permissions\\
Ensure that automatically created .pytest\_cache directories no longer use owner-only permissions after upgrading to pytest 8.2.1.\\
- Expectation: `.pytest\_cache` should receive normal directory permissions derived from the current process umask instead of owner-only access bits. \omittedtag{[omitted...]}\\
- Constraints: Permission handling must respect the active umask and avoid changing the public cache API shape. \omittedtag{[omitted...]}\\
\\
\#\# Additional Changes\\
- Ensure that users can verify the provenance of official pytest 8.2.1 source and wheel artifacts without changing pytest runtime behavior.\\

}
\end{AIBox}
\caption{L4 specification: problem statements with grounded expectations and constraints.}
\label{example:granularity-l4}
\end{example}
\clearpage

\begin{example}[ht]
\begin{AIBox}{L5: Full Synthesized Specification (\texttt{.md})}
{\ttfamily\scriptsize
\# Upgrade Specification from 8.2.0 to 8.2.1\\
\#\# New Features\\
\#\#\# Task 1 Add Python 3.13 compatibility and stricter pytester.makefile validation\\
The 8.2.1 update adds Python 3.13 support for runtime behaviors that depend on frame locals and makes invalid pytester file-creation calls fail explicitly when no extension is provided.\\
Ensure that the library works correctly on Python 3.13 for traceback and pytester workflows that depend on frame-local state, and ensure that invalid `Pytester.makefile()` calls with `ext=None` are rejected explicitly.\\
- Expectation: On Python 3.13, behavior that reads frame locals in traceback processing and in `pytester` hook-recording checks must continue to work even when frame locals are exposed through a proxy-style mapping rather than a plain dict. `Pytester.makefile()` must raise `TypeError` when `ext` is `None` instead of accepting that input.\\
- Constraints: Existing behavior for supported Python versions must remain intact. Recursion detection in tracebacks must still treat repeated frames with equivalent local-variable state as recursive. Valid `Pytester.makefile()` calls with a real extension string must continue to follow the existing extension rules, with only `None` being newly rejected.\\
- Acceptance Criteria:\\
\hspace*{2em}- When the library runs on Python 3.13 and traceback processing needs to compare local-variable state across repeated frames, it should still recognize recursion correctly.\\
\hspace*{2em}- When code uses `pytester` features that inspect recorded hook calls on Python 3.13, those checks should complete successfully without errors caused by non-dict frame locals.\\
\hspace*{2em}- When `Pytester.makefile()` is called with `ext=None`, it should raise `TypeError`.\\
\hspace*{2em}- When `Pytester.makefile()` is called with a valid extension string, it should continue to create files under the existing extension-validation rules.\\
- Behavioral Description:\\
\hspace*{2em}- On Python 3.13, traceback recursion detection continues to work when frame locals are no longer provided as a plain dict.\\
\hspace*{2em}- On Python 3.13, `pytester` hook-recording checks that evaluate against caller locals continue to work without failing because of the locals representation.\\
\hspace*{2em}- `Pytester.makefile()` now raises `TypeError` immediately when called with `ext=None`.\\
\\
\#\# Bug Fixes\\
\#\#\# Task 2 Avoid collection crashes from unreadable unselected directories\\
The update fixes a runtime collection failure in pytest when an unreadable directory exists in the project tree but is not part of the command-line selection. Instead of aborting collection with a PermissionError while checking whether that directory is a Python package, pytest should ignore that directory and continue collecting the explicitly requested tests.\\
Ensure that running pytest against selected test paths does not crash when unrelated directories in the project tree are unreadable.\\
- Expectation: During directory collection, \omittedtag{[omitted...]}\\
- Constraints: The fix is limited to PermissionError \omittedtag{[omitted...]}\\
- Acceptance Criteria: \omittedtag{[omitted...]}\\
- Behavioral Description: \omittedtag{[omitted...]}\\
\\
\#\#\# Task 3 Preserve system-exit behavior during pytest collection\\
\omittedtag{[omitted...]}\\
\#\#\# Task 4 Handle read-only squashfuse\_ll mounts without crashing\\
\omittedtag{[omitted...]}\\
\#\#\# Task 5 Restore default .pytest\_cache directory permissions\\
\omittedtag{[omitted...]}\\
\\
\#\# Additional Changes\\
- Ensure that users upgrading to pytest 8.2.1 can verify the provenance of official pytest distribution artifacts, including both source distributions and wheels, without changing pytest's runtime behavior.\\
}
\end{AIBox}
\caption{L5 specification: problem statements, grounded expectations and constraints, behaviors, and acceptance criteria.}
\label{example:granularity-l5}
\end{example}

\clearpage
\section{Limitations}
\label{app:limitations}
First, we do not conduct human quality assessment of the synthesized specifications.
Human quality assessment is prohibitively expensive even for a sampled subset, because it requires package-specific understanding of the release note, all linked issues or PRs, whole code changes, and the synthesized behavioral requirements.
It is too expensive at our scale of 155 version transitions and 1,660 grounded upgrade tasks, but we make the pipeline controllable, auditable, stable, and grounded through release-note and code-diff alignment, and schema constraints.
The stability check in Appendix~\ref{app:stability} shows that the matching agent remains stable even under the largest and most challenging transition.
Second, \methodname currently consists of only 12 Python package chains, so its coverage of other languages, build systems, and ecosystems remains limited.
Third, we evaluate eight models under three agent CLIs, with only two open-weight models, and future work should cover more agent frameworks and open-source systems.
Finally, we do not perform sampling-based repeated runs because long-horizon evaluation is costly, with a single chain costing up to around \$150 for strong models.

\section{Broader Impacts}
\label{app:broader}
\methodname supports research on reliable coding agents by evaluating continuous release-level software maintenance scenarios.
Agents must update existing codebases across consecutive package upgrades while resolving upgrade-related behaviors and avoiding regressions.
This can help identify reliability gaps before deploying coding agents in real development workflows.
Besides, our specification-granularity study in \S\ref{sec:res:granularity} provides a flexible setting for future research, enabling studies under both realistic upgrade guidance and more oracle-like conditions for probing upper-bound implementation capability and long-horizon reliability.

However, the main risks are overgeneralization and unsafe automation.
\methodname focuses on open-source Python packages with release metadata, code diffs, and executable test suites, so conclusions may not directly generalize to proprietary software, non-Python ecosystems, UI-heavy systems, or domains with poorly documented behavioral requirements.
Stronger coding agents may also enable large-scale, unsafe, or poorly reviewed code changes.
We mitigate these risks by framing \methodname as a controlled evaluation benchmark.
The dataset is built from public open-source resources, does not intentionally include private information, and evaluates agents in sandboxed Docker environments with restricted tool access.

\section{Declaration of LLM usage}
\label{app:declaration}
We used LLMs to revise paper drafts and assist with writing basic framework code, such as standard utility functions and implementation scaffolding.
All generated or assisted code was manually reviewed line by line by the authors before and after integration.

% \newpage
% \input{checklist.tex}

\end{document}